\documentclass[11pt]{article}
\usepackage{amssymb}
\usepackage{latexsym}
\usepackage{amsmath}
\usepackage{graphicx}

\usepackage{esdiff}
\usepackage{mathtools}

\def\vp{\varphi}
\def\b{\beta}
\def\ve{\varepsilon}
\def\al{\alpha}
\def\nb{\nabla}
\def\th{\theta}
\def\e{\eta}
\def\l{\lambda}
\def\k{\kappa}
\def\L{\Lambda}

\title{\LARGE{Static topological black hole with nonminimal derivative coupling and nonlinear electromagnetic field of Born-Infeld type}}
\author{M. M. Stetsko\footnote{E-mail: mstetsko@gmail.com}\
\\
  {\small Department for Theoretical Physics, Ivan Franko National University of Lviv,}\\
{\small 12 Drahomanov Str., Lviv, UA-79005, Ukraine
         }}

\begin{document}
\maketitle

{\abstract{We consider scalar-tensor gravity with nonminimal derivative coupling and Born-Infeld electromagnetic field which is minimally coupled to gravity. Since cosmological constant is taken into account it allowed us   not only derive static black hole with spherical horizon but also to obtain topological solutions with non-spherical horizons. The obtained metrics are thoroughly analyzed, namely for different distances and types of topology of horizon. To investigate singularities of the metrics Kretschmann scalar is used and it is shown that the character of singularity depends on the type of topology of horizon and dimension of space.  We also investigate black hole's thermodynamics, namely we obtain and examine black hole's temperature. To derive the first law of black hole's thermodynamics Wald's approach is applied. Nonetheless this approach is well established, there is ambiguity in definition of black hole's entropy which can be resolved just by virtue of some independent approach.}}
\section{Introduction} 
General Relativity is extremely successful theory which explains vast range of gravitational phenomena starting from the planetary motion and up to the evolution of the Universe \cite{Will_LRR2014,Berti_CQG2015}. Numerous observations in astrophysics go hand in hand with theoretically predicted values \cite{Abbott_PRL16-17}, but nonetheless there are still some open issues that give a chance for new theories which can be treated as generalization of Einsteinian General Relativity \cite{Clifton_PhysRept2012,Heisenberg}. The most puzzling questions which still remain unsolved are the problems related to the origin (or existence) of cosmological singularities, Dark Energy and Dark Matter issues, the problems of evolution of the early  Universe, for instance the problem of Inflation which is explained by approaches that take into account terms of the higher order of curvature \cite{Starobinsky_model}.

To find solutions of the mentioned above problems different sorts of modification of General Relativity were used. Among them we distinguish for example $F(R)$ theory, Lovelock Gravity, nonlocal modifications of the gravitational action, approaches which incorporate torsion tensor, namely Teleparallel Gravity, Scalar-Tensor Gravity. There are some interrelations between different of these approaches, but certainly any of them has its own peculiarities, more detailed description of the mentioned modifications of General Relativity is given in recent review \cite{Heisenberg}. 

Scalar-Tensor theories are probably the most conservative modification of General Relativity, because the departing point of all these theories is still General Relativity, but additional scalar fields are also included. The scalar fields can be coupled with gravitational degrees of freedom in different ways and due to the character of coupling we can classify all these theories. Important feature of the scalar-tensor theories is also coupling between the scalar and additional material fields which might be taken into consideration, but it should be pointed out here that usually it is supposed to be just minimal coupling between the material fields and gravity in order to obey the equivalence principle.

Among the various types of the scalar-tensor theories we would like to focus on the so-called Horndeski Gravity \cite{Horndeski_IJTP74}. One of the most remarkable features of this theory is related to the fact  that  equations of motion in Horndeski Gravity are of the second order, so it is free from ghost instabilities. Being beyond the scope in the gravity community for several decades it has been studied intensively since the time when relation between Horndeski Gravity and some scenarios in String Theory was established. Namely, this revival of interest in Horndeski Gravity is caused by the investigation of the so called Galileon Theories \cite{Nicolis_PRD09,Deffayet_PRD09} which are the scalar field theories that posses shift or Galilean symmetry, they are ghost free and those studies also renewed interest to DGP-model \cite{Dvali_PLB00}, which originally suffered from the ghost instabilities.  The equivalence between Horndeski and Galileon Theories was established \cite{Kobayashi_PTEP11}. Another approach which also gives rise to Galileon Theory takes its origin in Kaluza-Klein dimensional reduction procedure \cite{Acoleyen_PRD11, Charmousis_LNP15}. Some other approaches related to String Theory also give rise to Galileon-like models \cite{Cartier_PRD01,deRham_PRD11}. We note that multiscalar versions of Horndeski Gravity were also studied \cite{Deffayet_PRD10,Padilla_JHEP13,Charmousis_JHEP14,Ohashi_JHEP15} and  approaches that go beyond Horndeski Theory, but keeping their main attractive features were considered \cite{Zumala_PRD14,Gleyzes_PRL15,Langlois_JCAP16,Crisostomi_JCAP16,BenAchour_PRD16}.

Since its second revival Hondeski Gravity has been applied to vast range of problems in Cosmology and Physics of Black Holes. In particular, cosmological solutions and various cosmological scenarios were studied \cite{Sushkov_PRD09,Skugoreva_PRD13, Starobinsky_JCAP16}. Dynamics of Dark Energy/ Dark Matter models was investigated \cite{Granda_JCAP10,Gao_JCAP10,Sadjadi_PRD11}. Slow-roll inflation  mechanism without violation of unitarity bounds can be achieved in case of nonminimally coupled theories \cite{Germani_PRL10}. Various aspects of inflation were studied in theories with nonminimal derivative coupling, in particular reheating process during rapid oscillations and curvaton scenario were examined  \cite{Sadjadi_JCAP13,Dalianis_JCAP17,Feng_PLB14,Feng_PRD14,Qiu_EPJC17}. Effective field theory of Dark Energy was considered and its relation to Horndeski Gravity was established \cite{Kennedy_PRD17}. Predictions of Horndeski Gravity and constraints on various modifications of General Relativity due to modern experimentally obtained results were discussed  \cite{Heisenberg,Hou_EPJC17,Crisostomi_PRD18,Langlois_PRD18}.

A lot of attention has been paid to the investigation of compact objects such as neutron stars and black holes in Horndeski Gravity. Horndeski Gravity in its most general setting is known to have very complicated structure, thus the studies of the compact objects in this case are very difficult to perform. As a consequence, the investigation of the black holes and the neutron stars even in particular cases of Horndeski Gravity is of paramount importance. Rinaldi was the first who derived a static black hole solution in four dimensional case \cite{Rinaldi_PRD12}. Then static black holes' solutions were obtained and investigated for various dimensions and in more general setup of Horndeski Gravity \cite{Minamitsuji_PRD14,Anabalon_PRD14,Cisterna_PRD14,Kobayashi_PTEP14,Babichev_JHEP14,Bravo-Gaete_PRD14,Giribet_PRD15, Clement_CQG18}. Some attention has been paid to the examination of slowly rotating neutron stars and black holes \cite{Cisterna_PRD16,Maselli_PRD15,Stetsko_slr}. Some aspects of black hole thermodynamics were studied \cite{Feng_JHEP15,Feng_PRD16,Stetsko_PRD19}. Stability problem for different types of black holes, for instance with respect to odd-parity perturbations, was investigated  \cite{Cisterna_PRD15,Takahashi_PRD17,Tretyakova_CQG17,Ganguly_CQG18,Babichev_PRL18}. Important problem also related to stability of black holes, is the causal structure which was studied in \cite{Benkel_PRD18}. Boson and neutron stars in Horndeski Gravity were examined \cite{Cisterna_PRD16,Cisterna_PRD15,Cisterna_PRD15_2,Brihaye_PRD16,Verbin_PRD18}. Several other aspects of black hole's physics in Horndeski Theory such as the existence of hair \cite{Sotiriou_PRL14}, or influence of higher order terms over curvature \cite{Antoniou_PRL18} were investigated.

In this work we consider a particular case of Horndeski gravity, namely the theory with nonminimal derivative coupling and we also take into account nonlinear electromagnetic field of Born-Infeld type which is minimally coupled to gravity. Within this framework we find static solutions of field equations which represent  black holes. We point out here that charged static black hole solution in case of standard linear Maxwell field was studied in \cite{Cisterna_PRD14,Feng_PRD16}. In our previous work \cite{Stetsko_PRD19} we studied charged black hole, but with nonlinearity of other type, namely the so-called power-law nonlinearity. Some aspects of the present work were also examined in our recent work \cite{Stetsko_prep} where comparison between Born-Infeld and power-law cases was shown. Here we give deeper analysis of Born-Infeld case, with thorough investigation of the obtained solutions. We also study thermodynamics of the black holes, derive and examine their temperature and obtain the first law of black hole's thermodynamics.

We would like to point out here that Born-Infeld modification of gauge field Lagrangian \cite{Born_PRSA1934} and even its gravitational generalization, namely the so-called Eddington-inspired Born-Infeld gravity \cite{Banados_PRL10} are the subjects of active investigation for recent years \cite{BI_diff}. The revival of interest to the Born-Infeld theory was mainly inspired by studies in String Theory, namely it was shown \cite{Fradkin_PLB85} that Born-Infeld Lagrangian might be treated as a low energy effective theory which describes vector field coupled to the string ending on D-brane. Born-Infeld theory also appears in a bosonic sector of an effective theory which describes spontaneous breaking of supersymmetry \cite{Bagger_PRD97}. Apart of its importance for String Theory, the Born-Infeld theory possesses some attractive features which make this  theory rather exceptional among the other nonlinear modifications of a linear theory. Firstly, we would like to mention finiteness of a self-energy of a point particle, actually it was the aim which brought Born and Infeld to introduce such a modification of the linear Maxwell theory \cite{Born_PRSA1934}. Similarly to the linear vacuum Maxwell equation, vacuum Born-Infeld equations possess zero birefringence and have exceptional causal behaviour \cite{Boillat_JMP70}. Born-Infeld theory conserves helicity, that was shown in several independent ways \cite{Rosly_proc02,Boels_JHEP08,Novotny_PRD18}, namely it was argued to be a consequence of self-duality of the Born-Infeld theory. Born-Infeld theory was shown to be the only theory, except the standard Maxwell electrodynamics, with causal massless spin-$1$ propagation without supersymmetrization \cite{Deser_JPA80}. These peculiarities of the Born-Infeld theory were a good motivation to study Born-Infeld theory in different contexts \cite{BI_diff} and in particular they give solid arguments for us to consider Born-Infeld theory in the framework of the theory with nonminimal derivative coupling. We also note here that actually we consider not the general Born-Infeld theory, but rather its simplified form which nevertheless inherits the main features of the latter theory, this fact can be explained by complicated structures of both theories which we use here, namely the Scalar-Tensor theory of Gravity we use here and the Born-Infeld theory.

The organization of the present paper is the following: in the second section obtain and investigate the black holes solutions, it should be pointed out here that since we take into account cosmological constant apart of spherically symmetric black hole solutions with non-spherical topology of horizon are considered in the present work. Small subsection of the second section is devoted to the investigation of the gauge potential for the corresponding black holes. In the third section we investigate some aspects of thermodynamics of the black hole, namely the behaviour of its temperature is studied and it is compared with the results derived for the linear Maxwell field \cite{Feng_PRD16,Stetsko_PRD19}, in the second part of this section we obtain the first law of black hole's thermodynamics using Wald's approach. In the forth section we give some conclusions.

\section{Field equations and black hole's solution}
The action for the system we consider comprises of several terms, namely the standard Einstein-Hilbert part with cosmological constant, the terms describing the scalar field minimally and nonminimally coupled to gravity and finally the electromagnetic part, minimally coupled to gravity and which is represented by the action of Born-Infeld type. So this action can be written the form:
 \begin{equation}\label{action}
S=\frac{1}{16\pi}\int d^{n+1}x\sqrt{-g}\left( R-2\Lambda-\frac{1}{2}\left(\alpha g^{\mu\nu}-\eta G^{\mu\nu}\right)\partial_{\mu}\vp\partial_{\nu}\vp +4\b^2\left(1-\sqrt{1+\frac{F_{\mu\nu}F^{\mu\nu}}{2\b^2}}\right)\right)+S_{GHY}
\end{equation}
and here $g_{\mu\nu}$  is the metric tensor and $g=det{g_{\mu\nu}}$ denotes its determinant, $G_{\mu\nu}$ and $R$ are the Einstein tensor and Ricci scalar respectively, $\L$ is the cosmological constant and $\vp$ is the scalar field, nonminimally coupled to gravity, $F_{\mu\nu}=\partial_{\mu}A_{\nu}-\partial_{\nu}A_{\mu}$ is the Maxwell field tensor and $\b$ denotes the Born-Infeld coupling constant. It should be noted that when $\b\rightarrow\infty$ the Born-Infeld term gets transformed into standard linear Maxwell term, we also remark that the parameter $\beta$ is supposed to be positive to obtain reasonable physical results which are in agreement with linear field case. Finally, $S_{GHY}$ denotes boundary Gibbons-Hawking-York term, which is taken into account to have the variational problem well defined, and in case of the theory with nonminimal derivative coupling it can be represented in the form:
\begin{equation}\label{GHY_nm}
S_{GHY}=\frac{1}{8\pi}\int d^nx\sqrt{|h|}\left(K+\frac{\e}{4}\left[\nb^{\mu}\vp\nb^{\nu}\vp K_{\mu\nu}+(n^{\mu}n^{\nu}\nb_{\mu}\vp \nb_{\nu}\vp+(\nb\vp)^2)K\right]\right),
\end{equation}
and here $h$ denotes the determinant of the boundary metric $h_{\mu\nu}$, $K_{\mu\nu}$ and $K$ are the extrinsic curvature tensor and  its trace respectively, $n_{\mu}$ is the normal to the boundary hypersurface. As it is known Gibbons-Hawking-York term does not affect on the equations of motion in the bulk.

 As we have noted in the introduction, the gauge field part of the Lagrangian in the action (\ref{action}) should be treated as a particular case of a general Born-Infeld Lagrangian which can be cast in the form:
\begin{equation}\label{BI_lagr}
 {\cal L}_{BI}=\sqrt{-det(\b g_{\mu\nu})}-\sqrt{-det(\b g_{\mu\nu}+F_{\mu\nu})},
\end{equation}
where $g_{\mu\nu}$ is the metric tensor, $F_{\mu\nu}$ is the gauge field tensor introduced above and $\b$ again denotes the Born-Infeld coupling constant. If one would like to take into account the lowest order contribution over $\b$, but keeping the main features of the general Born-Infeld theory, the Lagrangian (\ref{BI_lagr}) can be rewritten in the form:
\begin{equation}\label{BI_simpl}
 {\cal L}_{BI}\simeq\b^{(n+1)/2}\sqrt{-g}\left(1-\sqrt{1+\frac{F_{\mu\nu}F^{\mu\nu}}{2\b^2}}\right).
\end{equation}
The latter relation coincides with the Lagrangian for the gauge field in the action (\ref{action}) with respect to the constant factor $\b^{(n-3)/2}$. We also point out here that under assumption about the form of the gauge potential $A_{\mu}$ or consequently the field tensor $F_{\mu\nu}$ which we use below, the right hand side of the latter relation is exactly the same as in the case of general Born-Infeld theory (\ref{BI_lagr}), this fact can be verified easily.

The principle of the least action, applied to the action (\ref{action}) gives rise to equations of motion for the system. Namely, for the gravitational field the equations of motion can be written in the form:
\begin{equation}\label{eom}
G_{\mu\nu}+\Lambda g_{\mu\nu}=\frac{1}{2}(\alpha T^{(1)}_{\mu\nu}+\eta T^{(2)}_{\mu\nu})+T^{(3)}_{\mu\nu}
\end{equation}
where the following notations are used:
\begin{equation}\label{scal_min}
T^{(1)}_{\mu\nu}=\nb_{\mu}\vp\nb_{\nu}\vp-\frac{1}{2}g_{\mu\nu}\nb^{\lambda}\vp\nb_{\lambda}\vp,
\end{equation}
\begin{eqnarray}\label{scal_nm}
\nonumber T^{(2)}_{\mu\nu}=\frac{1}{2}\nb_{\mu}\vp\nb_{\nu}\vp R-2\nb^{\lambda}\vp\nb_{\nu}\vp R_{\lambda\mu}+\frac{1}{2}\nb^{\lambda}\vp\nb_{\lambda}\vp G_{\mu\nu}-g_{\mu\nu}\left(-\frac{1}{2}\nb_{\lambda}\nb_{\kappa}\vp\nb^{\lambda}\nb^{\kappa}\vp\right.\\\left.+\frac{1}{2}(\nb^2\vp)^2-R_{\lambda\kappa}\nb^{\lambda}\vp\nb^{\kappa}\vp\right)
-\nb_{\mu}\nb^{\lambda}\vp\nb_{\nu}\nb_{\lambda}\vp+
\nb_{\mu}\nb_{\nu}\vp\nb^2\vp-R_{\lambda\mu\kappa\nu}\nb^{\lambda}\vp\nb^{\kappa}\vp
\end{eqnarray}
\begin{equation}\label{max_tr_nlin}
T^{(3)}_{\mu\nu}=2\b^2g_{\mu\nu}\left(1-\sqrt{1+\frac{F_{\k\l}F^{\k\l}}{2\b^2}}\right)+\frac{2F_{\mu\rho}{F_{\nu}}^{\rho}}{\sqrt{1+\frac{F_{\k\l}F^{\k\l}}{2\b^2}}}
\end{equation}  
It is worth being remarked that the term $T^{(1)}_{\mu\nu}$ is the standard form of stress-energy tensor for a minimally coupled scalar field  and $T^{(2)}_{\mu\nu}$ corresponds to the stress-energy tensor of the nonminimally coupled part. Finally, the term $T^{(3)}_{\mu\nu}$ denotes the stress-energy tensor for the electromagnetic field given by the Born-Infeld action. Varying the action (\ref{action}) with respect to the scalar field $\vp$ one arrives at the following equation:
\begin{equation}\label{scal_f_eq}
(\alpha g_{\mu\nu}-\eta G_{\mu\nu})\nb^{\mu}\nb^{\nu}\vp=0.
\end{equation}
Taking the variation of the action (\ref{action}) with respect to the gauge potential $A_{\mu}$ we obtain the equations for the gauge field:
\begin{equation}\label{Maxwell_eq}
\nb_{\mu}\left(\frac{F^{\mu\nu}}{\sqrt{1+\frac{F_{\k\l}F^{\k\l}}{2\b^2}}}\right)=0.
\end{equation}
It should be noted that in the limit $\b\rightarrow\infty$ standard Maxwell equations are recovered.

Since the action (\ref{action}) includes the cosmological constant $\L$ and it means that topological solutions might exist. Here we will consider  static topological solutions and consequently suppose that the metric takes the form:
\begin{equation}\label{metric}
ds^2=-U(r)dt^2+W(r)dr^2+r^2d\Omega^{2(\ve)}_{(n-1)},
\end{equation}
where $d\Omega^{2(\ve)}_{n-1}$ is the line element of a $n-1$--dimensional hypersurface of a constant curvature which can be represented in the form: 
\begin{eqnarray}
d\Omega^{2(\ve)}_{(n-1)}=
\begin{cases}
d\th^2+\sin^2{\th}d\Omega^2_{(n-2)}, \quad \ve=1,\\
d\th^2+{\th}^2 d\Omega^2_{(n-2)},\quad \ve=0,\\
d\th^2+\sinh^2{\th}d\Omega^2_{(n-2)},\quad \ve=-1,
\end{cases}
\end{eqnarray}
and here $d\Omega^2_{(n-2)}$ is the line element of a $n-2$--dimensional hypersphere. Thus, the expression $d\Omega^{2(\ve)}_{(n-1)}$ represents the line element on a hypersurface of positive, null or negative curvature for given values of parameter $\ve$. We point out here that in the present work we consider dimensions $n\geqslant 3$.

As it has been mentioned above we are going to obtain the static solutions and it means that we can choose the gauge field form as follows: $A=A_{0}(r)dt$ (only a scalar component of electromagnetic potential is nonzero). Taking the chosen form of gauge field and the metric (\ref{metric}) into account we solve the equations (\ref{Maxwell_eq}) and obtain the electromagnetic field tensor of the form:
\begin{equation}\label{EM_field}
F_{rt}=-\frac{q\b}{\sqrt{q^2+\b^2r^{2(n-1)}}}\sqrt{UW},
\end{equation}
where $q$ is an integration constant related to the black hole's charge. We note here that the behaviour of the electromagnetic field depends on the product of metric functions $UW$ which is a function of the radial coordinate $r$.

Taking into consideration the explicit form of the metric (\ref{metric}) and integrating the equation (\ref{scal_f_eq}) for a once we arrive at the relation:
\begin{equation} 
\sqrt{\frac{U}{W}}r^{n-1}\left[\al-\e\frac{(n-1)}{2rW}\left(\frac{U\rq{}}{U}-\frac{(n-2)}{r}(\ve W-1)\right)\right]\vp'=C
\end{equation}
In the following we assume that obtained in the latter relation constant $C$ is equal to zero and it simplifies the procedure of solving of the field equations (\ref{eom}). The imposed condition on the constant $C$ is equivalent to the assumption about the relation between a component of the metric tensor $g_{\mu\nu}$ and Einstein tensor $G_{\mu\nu}$, namely:
\begin{equation}\label{cond_nm}
\al g_{rr}-\e G_{rr}=0
\end{equation}
We note that the relation (\ref{cond_nm}) was used in most papers where black holes in the theory with nonminimal derivative coupling were examined, in particular it appears in \cite{Rinaldi_PRD12,Minamitsuji_PRD14,Feng_JHEP15,Feng_PRD16,Stetsko_PRD19}.

Taking into account the evident form for the metric (\ref{metric}) and the obtained above expression for electromagnetic field (\ref{EM_field}) we can represent the equations (\ref{eom}) in the form:
\begin{eqnarray}\label{G_0}
\nonumber \frac{(n-1)}{2rW}\left(\frac{W\rq{}}{W}+\frac{(n-2)}{r}(\ve W-1)\right)\left(1+\frac{3}{4}\e\frac{(\vp\rq{})^2}{W}\right)-\L=\frac{\al}{4W}(\vp\rq{})^2+\\\frac{\e}{2}\left(\frac{(n-1)}{rW^2}{\vp''}\vp\rq{}+\frac{(n-1)(n-2)}{r^2W^2}(\vp\rq{})^2\left(\ve W-\frac{1}{2}\right)\right)-2\b^2+\frac{2\b}{r^{n-1}}\sqrt{q^2+\b^2 r^{2(n-1)}};
\end{eqnarray}
\begin{eqnarray}\label{G_1}
\nonumber \frac{(n-1)}{2rW}\left(\frac{U\rq{}}{U}-\frac{(n-2)}{r}(\ve W-1)\right)\left(1+\frac{3}{4}\e\frac{(\vp\rq{})^2}{W}\right)+\L=\\\frac{\al}{4W}(\vp\rq{})^2-\frac{\e}{2}\ve\frac{(n-1)(n-2)}{2r^2W}(\vp\rq{})^2+2\b^2-\frac{2\b}{r^{n-1}}\sqrt{q^2+\b^2 r^{2(n-1)}};
\end{eqnarray}

\begin{eqnarray}\label{G_2}
\nonumber\left[\frac{1}{2UW}\left(U\rq{}\rq{}-\frac{(U\rq{})^2}{2U}-\frac{U\rq{}W\rq{}}{2W}\right)+\frac{n-2}{2rW}\left(\frac{U\rq{}}{U}-\frac{W\rq{}}{W}\right)-\frac{(n-2)(n-3)}{2r^2W}(\ve W-1)\right]\times \\\nonumber\left(1+\frac{\e}{4}\frac{(\vp\rq{})^2}{W}\right)+\L=-\frac{\al}{4W}(\vp\rq{})^2-\frac{\e}{2W^2}\vp\rq{}\rq{}\vp\rq{}\left(\frac{U\rq{}}{2U}+\frac{n-1}{r}\right)+\\\frac{\e}{2}\frac{(\vp\rq{})^2}{W}\left(\frac{U\rq{}W\rq{}}{4UW^2}+\frac{(n-2)W\rq{}}{2rW^2}-\ve\frac{(n-2)(n-3)}{2r^2}\right)+2\b^2\left(1-\frac{\b r^{n-1}}{\sqrt{q^2+\b^2r^{2(n-1)}}}\right).
\end{eqnarray} 
It is worth noting that we have four equations, namely (\ref{cond_nm}), (\ref{G_0}), (\ref{G_1}) and (\ref{G_2}) for three unknown functions $U(r)$, $W(r)$ and $\vp'$, it means the we might choose three of them to find the unknown functions and the fourth equation should be satisfied as an identity for the obtained solution. From the reason of simplicity we take the equations (\ref{cond_nm}), (\ref{G_0}) and (\ref{G_1}). Having used the equations (\ref{G_0}) and (\ref{G_1}) we can write:
\begin{equation}\label{fi_der_2}
(\vp')^2=-\frac{4r^2W}{\e(2\al r^2+\ve\e(n-1)(n-2))}\left(\al+\L\e-2\b^2\e+2\b\e r^{1-n}\sqrt{q^2+\b^2r^{2(n-1)}}\right);
\end{equation}
\begin{equation}\label{UW_prod}
UW=\frac{\left((\al-\L\e+2\b^2\e)r^2+\ve\e(n-1)(n-2)-2\b\e r^{3-n}\sqrt{q^2+\b^2r^{2(n-1)}}\right)^2}{(2\al r^2+\ve\e(n-1)(n-2))^2}.
\end{equation}
We point out here that the right hand side of the relation (\ref{fi_der_2}) has to take positive values outside black hole's horizon. This condition leads to some restrictions on the parameters of coupling $\al$, $\e$, cosmological constant $\L$, charge parameter $q$ and Born-Infeld parameter $\b$. Having supposed that the coupling parameters $\al$ and $\e$ are positive (and assuming that the expression $2\al r^2+\ve\e(n-1)(n-2)$ is positive in the outer domain) one arrives at the conclusion that the expression $\al+\L\e-2\b^2\e+2\b\e r^{1-n}\sqrt{q^2+\b^2r^{2(n-1)}}$ should be negative outside the horizon, what can be achieved if the cosmological constant $\L$ is negative. We point out here that similar condition on the cosmological constant was imposed in case of neutral \cite{Stetsko_slr} and charged \cite{Stetsko_PRD19} black holes in Horndeski gravity. Since the metric function $W(r)$ diverges on the horizon changing its sign while crossing  the horizon, the same is true for the function $(\vp')^2$, unless the expression $\al+\L\e-2\b^2\e+2\b\e r^{1-n}\sqrt{q^2+\b^2r^{2(n-1)}}$ also changes it sign on the horizon. It leads to the consequence that in the inner domain $\vp'$ becomes purely imaginary (phantom-like behaviour), similar situation takes place in case of neutral and charged black holes \cite{Stetsko_slr, Stetsko_PRD19}. We point out there that the kinetic energy of the scalar field $K=\nb_{\mu}\vp\nb^{\mu}\vp$ is finite at the horizon and is positive in the inner domain up to the moment when the expression $\al+\L\e-2\b^2\e+2\b\e r^{1-n}\sqrt{q^2+\b^2r^{2(n-1)}}$  changes it sign. Similar analysis performed under assumption that $\e<0$ while $\al>0$ shows that in this case the cosmological constant $\L$ should be taken negative as well. Here we also note that if $\al+\L\e-2\b^2\e=0$, the right hand side of the relation (\ref{fi_der_2}) becomes negative in the outer domain and it is not acceptable from the physical point of view. If one imposes the condition $\al-\L\e+2\b^2\e=0$ to provide positivity of the right hand side of the relation (\ref{fi_der_2}) the parameter $\e$ should take negative values,  this condition allows to obtain the following relations for the metric function in a bit simpler form, but it does not change the character of asymptotic behaviour of the metric function $U(r)$ for small and large distances in comparison with the general case as it might be shown from the relations to be obtained. The right hand side of the relation (\ref{UW_prod}) is always nonnegative and it demonstrates that the metric functions $U(r)$ and $W(r)$ always have the same sign as it should be for a black hole's solution. It is worth emphasizing that in the limit $\beta\rightarrow\infty$ the relations (\ref{fi_der_2}), (\ref{UW_prod}) as well as the relations for the metric function $U(r)$  that will be obtained below, are reduced to corresponding relations derived for linear field \cite{Cisterna_PRD14,Feng_PRD16,Stetsko_PRD19}.

The metric function $U(r)$ can be written as follows:
\begin{eqnarray}\label{U_int}
\nonumber U(r)=\ve-\frac{\mu}{r^{n-2}}-\frac{2(\L-2\b^2)}{n(n-1)}r^2-\frac{2\b(\al-\L\e+2\b^2\e)}{\al(n-1)r^{n-2}}\int\sqrt{q^2+\b^2r^{2(n-1)}}dr+\frac{(\al+\L\e-2\b^2\e)^2}{2(n-1)\al\e r^{n-2}}\times\\\int\frac{r^{n+1}}{r^2+d^2}dr-\frac{2\b(\al+\L\e-2\b^2\e)d^2}{\al(n-1) r^{n-2}}\int\frac{\sqrt{q^2+\b^2r^{2(n-1)}}}{r^2+d^2}dr+\frac{2\b^2\e}{\al(n-1)r^{n-2}}\int\frac{r^{3-n}(q^2+\b^2r^{2(n-1)})}{r^2+d^2}dr,
\end{eqnarray}
where $d^2=\ve\e(n-1)(n-2)/2\al$ and we point out that in the following relations we assume that $d^2>0$. It should be noted that the second and the fourth integrals in the relation (\ref{U_int}) have some differences for odd and even dimensions of space $n$ and the first and third ones cannot be expressed in terms of elementary functions.  Namely the first integral when $q^2r^{2(1-n)}/\b^2<1$ can be represented as follows \cite{Bateman}:
\begin{equation}\label{hyp_large}
\int\sqrt{q^2+\b^2r^{2(n-1)}}dr=\frac{\b}{n}r^n{_{2}F_{1}}\left(-\frac{1}{2},\frac{n}{2(1-n)};\frac{2-n}{2(1-n)};-\frac{q^2}{\b^2}r^{2(1-n)}\right)
\end{equation}
and we additionally stress here that the right hand side of the written above relation is valid for large distances ($r^{2(n-1)}>q^2/\b^2$), whereas for small distances $r^{2(n-1)}<q^2/\b^2$ we can write:
\begin{equation}\label{hyp_small}
\int\sqrt{q^2+\b^2r^{2(n-1)}}dr=qr{_{2}F_{1}}\left(-\frac{1}{2},\frac{1}{2(n-1)};\frac{2n-1}{2(n-1)};-\frac{\b^2}{q^2}r^{2(n-1)}\right).
\end{equation}
The third integral in (\ref{U_int}) needs also special care and due to its importance we pay attention to it. We point out here that the result of calculation of the mentioned integral depends on the parity of $n$ and similarly to the written above relations (\ref{hyp_large}) and (\ref{hyp_small}) its evident form is defined by the condition whether $r^{2(n-1)}>q^2/\b^2$ or $r^{2(n-1)}<q^2/\b^2$ , but it also depends on the relation between $r$ and $d$ (takes a bit different form for $r>d$ and $r<d$). Namely, for odd $n$ when $r^{2(n-1)}>q^2/\b^2$ and $r>d$ (large distances) this integral can be represented in the form:
\begin{equation}\label{int_1}
\int\frac{\sqrt{q^2+\b^2r^{2(n-1)}}}{r^2+d^2}dr=\b r^{n-2}\sum^{+\infty}_{j=0}\frac{(-1)^j}{n-2(j+1)}\left(\frac{d}{r}\right)^{2j}{_{2}}F_{1}\left(-\frac{1}{2},\frac{n-2(j+1)}{2(1-n)};\frac{n+2j}{2(n-1)};-\frac{q^2}{\b^2}r^{2(1-n)}\right),
\end{equation}
if $r<d$, but still $r^{2(n-1)}>q^2/\b^2$ we can represent the latter integral in the following form:
\begin{equation}\label{int_2}
\int\frac{\sqrt{q^2+\b^2r^{2(n-1)}}}{r^2+d^2}dr=\frac{\b r^{n}}{d^2}\sum^{+\infty}_{j=0}\frac{(-1)^j}{n+2j}\left(\frac{r}{d}\right)^{2j}{_{2}}F_{1}\left(-\frac{1}{2},\frac{n+2j}{2(1-n)};\frac{n-2(j+1)}{2(n-1)};-\frac{q^2}{\b^2}r^{2(1-n)}\right).
\end{equation}
There is no difficulties with the latter integral when two other options for distance are given, namely when $r^{2(n-1)}<q^2/\b^2$ and $r<d$ or $r>d$. In contrast to odd $n$, the latter integral has some subtleties when $n$ is even. If $r^{2(n-1)}>q^2/\b^2$ and $r>d$ it can be written in the form:
\begin{eqnarray}\label{int_1e}
\nonumber\int\frac{\sqrt{q^2+\b^2r^{2(n-1)}}}{r^2+d^2}dr=\b r^{n-2}\mathop{\sum^{+\infty}_{j=0}}_{ j\neq \frac{n}{2}-1}\frac{(-1)^j}{n-2(j+1)}\left(\frac{d}{r}\right)^{2j}{_{2}}F_{1}\left(-\frac{1}{2},\frac{n-2(j+1)}{2(1-n)};\frac{n+2j}{2(n-1)};-\frac{q^2}{\b^2}r^{2(1-n)}\right)\\+(-1)^{\frac{n}{2}-1}\b d^{n-2}\left(\sum^{+\infty}_{l=1}\frac{(-1)^l}{l!}\left(-\frac{1}{2}\right)_{l}\left(\frac{q}{\b}\right)^{2l}\frac{r^{2(1-n)l}}{2(1-n)l}+\ln{\left(\frac{r}{d}\right)}\right),
\end{eqnarray}
where $(a)_l$ denotes the Pochhammer symbol of $a$. When $r<d$,  but still $r^{2(n-1)}>q^2/\b^2$  we can write:
\begin{eqnarray}\label{int_2e}
\nonumber\int\frac{\sqrt{q^2+\b^2r^{2(n-1)}}}{r^2+d^2}dr=\b\frac{r^n}{d^2}\mathop{\sum^{+\infty}_{j=0,}\sum^{+\infty}_{l=0}}_{j\neq (n-1)l-\frac{n}{2}}\frac{(-1)^{j+l}}{l!(n+2j+2(1-n)l)}\left(-\frac{1}{2}\right)_l\left(\frac{r}{d}\right)^{2l}\left(\frac{q}{\b}r^{1-n}\right)^{2l}\\+\b d^{n-2}\sum^{+\infty}_{l=1}\frac{(-1)^{nl-\frac{n}{2}}}{l!}\left(-\frac{1}{2}\right)_l\left(\frac{q}{\b}d^{1-n}\right)^{2l}\ln{\left(\frac{r}{d}\right)}.
\end{eqnarray}
If $r^{2(n-1)}<q^2/\b^2$ and $r<d$ or $r>d$ one can also write the evident form for the above integral. It should be pointed out that the main difference between odd and even $n$ cases for given above integral is the presence of logarithmic terms $\sim\ln(r/d)$ for even $n$, whereas for odd $n$ such a term does not appear.

 Considering large distance case when $r^{2(n-1)}>q^2/\b^2$ and $r>d$  and taking into account the written above relations (\ref{hyp_large}) and (\ref{int_1}) or (\ref{int_1e}) we can write the explicit form for the metric function $U$, namely for odd $n$ it takes the form:
\begin{eqnarray}\label{U_odd}
\nonumber U(r)=\ve-\frac{\mu}{r^{n-2}}-\frac{2(\L-2\b^2)}{n(n-1)}r^2-\frac{2\b^2(\al-\L\e+2\b^2\e)}{\al n(n-1)}r^2{_{2}F_{1}}\left(-\frac{1}{2},\frac{n}{2(1-n)};\frac{n-2}{2(n-1)};-\frac{q^2}{\b^2}r^{2(1-n)}\right)+\\\nonumber\frac{(\al+\L\e-2\b^2\e)^2}{2\al\e(n-1)}\left[\sum^{(n-1)/2}_{j=0}(-1)^j\frac{d^{2j}r^{2(1-j)}}{n-2j}+(-1)^{\frac{n+1}{2}}\frac{d^n}{r^{n-2}}\arctan{\left(\frac{r}{d}\right)}\right]-\frac{2\b^2(\al+\L\e-2\b^2\e)d^2}{\al(n-1)}\times\\\nonumber\sum^{+\infty}_{j=0}\frac{(-1)^j}{n-2(j+1)}\left(\frac{d}{r}\right)^{2j}{_{2}F_{1}}\left(-\frac{1}{2},\frac{n-2(j+1)}{2(1-n)};\frac{n+2j}{2(n-1)};-\frac{q^2}{\b^2}r^{2(1-n)}\right)+\frac{2\b^2\e}{\al(n-1)}\times\\\left[q^2\sum^{(n-5)/2}_{j=0}\frac{(-1)^jr^{6-2n+2j}}{(4-n+2j)d^{2(j+1)}}+\frac{(-1)^{\frac{n-3}{2}}}{r^{n-2}}\left(\frac{q^2}{d^{n-2}}+\b^2d^n\right)\arctan\left(\frac{r}{d}\right)+\b^2\sum^{(n-1)/2}_{j=0}(-1)^j\frac{d^{2j}r^{2(1-j)}}{n-2j}\right];
\end{eqnarray}
and for even $n$ we obtain:
\begin{eqnarray}\label{U_even}
\nonumber U(r)=\ve-\frac{\mu}{r^{n-2}}-\frac{2(\L-2\b^2)}{n(n-1)}r^2-\frac{2\b^2(\al-\L\e+2\b^2\e)}{\al n(n-1)}r^2{_{2}F_{1}}\left(-\frac{1}{2},\frac{n}{2(1-n)};\frac{n-2}{2(n-1)};-\frac{q^2}{\b^2}r^{2(1-n)}\right)+\\\nonumber\frac{(\al+\L\e-2\b^2\e)^2}{2\al\e(n-1)}\left[\sum^{n/2-1}_{j=0}(-1)^j\frac{d^{2j}r^{2(1-j)}}{n-2j}+(-1)^{\frac{n}{2}}\frac{d^n}{2r^{n-2}}\ln{\left(\frac{r^2}{d^2}+1\right)}\right]-\frac{2\b^2(\al+\L\e-2\b^2\e)d^2}{\al(n-1)}\times\\\nonumber\left[
\mathop{\sum^{+\infty}_{j=0}}_{j\neq\frac{n}{2}-1}\frac{(-1)^j}{n-2(j+1)}\left(\frac{d}{r}\right)^{2j}{_{2}F_{1}}\left(-\frac{1}{2},\frac{n-2(j+1)}{2(1-n)};-\frac{n+2j}{2(1-n)};-\frac{q^2}{\b^2}r^{2(1-n)}\right)+(-1)^{\frac{n}{2}}\frac{d^{n-2}}{r^{n-2}}\times\right.\\\nonumber\left.\left(\sum^{+\infty}_{j=1}\frac{(-1)^j}{j!}\left(-\frac{1}{2}\right)_{j}\left(\frac{q}{\b}\right)^{2j}\frac{r^{2(1-n)j}}{2(n-1)j}-\ln{\left(\frac{r}{d}\right)}\right)\right]+\frac{2\b^2\e}{\al(n-1)}\left[q^2\sum^{(n-6)/2}_{j=0}\frac{(-1)^jr^{6-2n+2j}}{(4-n+2j)d^{2(j+1)}}+\right.\\\left.\frac{(-1)^{\frac{n-2}{2}}}{2r^{n-2}}\left(\frac{q^2}{d^{n-2}}\ln\left(1+\frac{d^2}{r^2}\right)-\b^2d^n\ln\left(1+\frac{r^2}{d^2}\right)\right)+\b^2\sum^{n/2-1}_{j=0}(-1)^j\frac{d^{2j}r^{2(1-j)}}{n-2j}\right].
\end{eqnarray}
It should be stressed that the infinite sums in the written above relations (\ref{U_odd}) and (\ref{U_even}) are convergent when $r>d$ and $r^{2(n-1)}>q^2/\b^2$ (which also takes place for large distances), but there is no difficulty in writing the evident form for the metric function $U(r)$ when $r^{2(n-1)}>q^2/\b^2$ and $d<r$ or for other two possible options for the distance (small distances). As we have noted above, the condition $d^2>0$ is imposed after the integral form for the metric function $U(r)$ is written, we point out here that solutions with $d^2<0$ might be studied, but similarly as it was shown for neutral black hole \cite{Stetsko_slr} or power-law field \cite{Stetsko_PRD19} the corresponding solutions do not represent a black hole.    

For the flat horizon case ($\ve=0$) the metric function $U(r)$ can be written in a simpler form:
\begin{eqnarray}\label{U_flat}
\nonumber U(r)=-\frac{\mu}{r^{n-2}}+\frac{(\al-\L\e+2\b^2\e)^2+4\b^4\e^2}{2\al\e n(n-1)}r^2-\frac{2\b^2(\al-\L\e+2\b^2\e)}{\al n(n-1)}r^2\times\\{_{2}F_{1}}\left(-\frac{1}{2},\frac{n}{2(1-n)};\frac{2-n}{2(1-n)};-\frac{q^2}{\b^2}r^{2(1-n)}\right)-\frac{2\b^2\e q^2}{\al(n-1)(n-2)}r^{2(2-n)}. 
\end{eqnarray}
We note that the written above relation takes place for large $r$, for small $r$ instead of the relation (\ref{hyp_large}) the relation (\ref{hyp_small}) should be used, completely in the same way as it was done for the previously considered cases when $\ve\neq 0$.

Taking into account all the obtained expressions for the metric function $U$ and the product $UW$ we might investigate their behaviour for some specific values of $r$. Firstly, we examine the function $U(r)$  for small values of $r$. Having used the relations (\ref{U_odd}) and (\ref{U_even}) rewritten for small distances and considering only the leading terms we write:
\begin{equation}\label{u_r0_c}
U(r)\simeq\frac{4\b^2q^2}{\ve(n-1)^2(n-2)(4-n)}r^{2(3-n)}
\end{equation}
and here we omit all the subleading terms for small $r$. It should be  pointed out that the relation (\ref{u_r0_c}) is valid when $n\geqslant 5$. If $n=3$ the leading term is of the form:
\begin{equation}\label{u_r0_3}
U(r)\simeq-\frac{\mu}{r},
\end{equation}
in this case the dominant term for small distances is the same as for the nonminimally coupled theory without any electromagnetic field \cite{Stetsko_slr}. This fact can be explained by nonsingular behaviour of the electromagnetic potential at the origin of coordinates for the case $n=3$ and what is not true for higher dimensions. When $n=4$ the leading term can be written in the following form:
\begin{equation}\label{u_r0_c4}
U(r)\simeq -\frac{\b^2q^2}{9\ve r^2}\ln{\left(1+\frac{d^2}{r^2}\right)},
\end{equation}
as it is easy to conclude for $n=4$ the leading term (\ref{u_r0_c4}) is caused by the term of the same origin as in the relation (\ref{u_r0_c}), but due to different powers of $r$ under integral from which they are derived they can have either power-law or logarithmic dependences. The product of the metric functions $UW$ for small $r$ and $n\geqslant 4$ takes the form:
\begin{equation}\label{UW_r0_c}
UW\simeq\frac{4\b^2q^2}{(n-1)^2(n-2)^2}r^{2(3-n)},
\end{equation}
whereas for $n=3$ the product $UW\simeq\left(1-\frac{\b q}{\ve}\right)^2$ and when $q=0$ it goes to the limit that is typical for static black holes in standard General Relativity. We note that singular behaviour of the product $UW$ for small $r$ takes place for the black holes with linear and power-law nonlinear electromagnetic filed in the theory with nonminimal derivative coupling \cite{Feng_PRD16,Stetsko_PRD19}.

In the above analysis we have not considered the solution with flat horizon surface (\ref{U_flat}). As it is easy to see, the leading term for small distances takes the form as follows:
\begin{equation}\label{u_r0_f}
U(r)\simeq-\frac{2\b^2\e q^2}{\al(n-1)(n-2)}r^{2(2-n)},
\end{equation}
here we note that the latter equation takes place for all $n\geqslant 3$. Having compared the relations (\ref{u_r0_c}) and (\ref{u_r0_f}) one can conclude that for the case of flat horizon ($\ve=0$) the singularity of the metric function $U(r)$ when $r\rightarrow 0$ is stronger than for nonflat horizon surface ($\ve=\pm 1$). We also remark that for small distances for $\ve=0$ and $\ve=1$ the leading term is negative and for $\ve=-1$ it might be positive.

For large $r$ the metric functions (\ref{U_odd}), (\ref{U_even}) and (\ref{U_flat}) have similar dependence given by the leading term of the  form:
\begin{equation}
U\simeq \frac{(\al-\L\e)^2}{2n(n-1)\al\e}r^2,
\end{equation} 
and here similarly to small distances we do not write subleading terms. It is easy to see that for large $r$ the metric function is always of anti-de Sitter type. The product of the metric functions for large $r$ is as follows $UW\simeq (\al-\L\e)^2/4\al^2$.

We also consider the regime of large $\e$, namely when the terms which correspond to the minimal coupling (terms related to the parameter $\al$) are supposed to be considerably smaller than the terms which appear due to the presence of nonminimal coupling (terms related to $\eta$). The metric function $U(r)$ takes the form:
\begin{eqnarray}\label{U_large_e}
\nonumber U(r)\simeq\ve-\frac{\mu}{r^{n-2}}+\frac{4\b^2 q^2}{\ve(n-1)^2(n-2)(4-n)}r^{2(3-n)}-\frac{2(\L-2\b^2)}{n(n-1)}r^2+\\\nonumber\frac{(\L-2\b^2)^2+4\b^4}{\ve(n-1)^2(n^2-4)}r^4-\frac{4\b^2}{n(n-1)}r^2{_{2}F_{1}}\left(-\frac{1}{2},\frac{n}{2(1-n)};\frac{2-n}{2(1-n)};-\frac{q^2}{\b^2}r^{2(1-n)}\right)+\\\frac{4\b^2(\L-2\b^2)r^4}{\ve(n-1)^2(n^2-4)}{_{2}F_{1}}\left(-\frac{1}{2},\frac{n+2}{2(1-n)};\frac{4-n}{2(1-n)};-\frac{q^2}{\b^2}r^{2(1-n)}\right)+{\cal O}\left(\frac{1}{\e}\right).
\end{eqnarray}
Having compared the expression (\ref{U_large_e}) with (\ref{U_flat}) one can conclude that for the cases of nonflat horizon ($\ve=\pm 1$) the written above relation does not contain the terms proportional to $\e$ whereas the relation (\ref{U_flat}) does, so they have completely different behavior in this regime. We also remark that similar situation took place for nonlinear electromagnetic field with power-law dependence \cite{Stetsko_PRD19}. The product of the metric functions in regime of large $\e$ for nonflat topology of horizon takes the form:
\begin{equation}
UW\simeq\left(1-\frac{\L-2\b^2}{\ve(n-1)(n-2)}r^2-\frac{2\b r^{3-n}}{\ve(n-1)(n-2)}\sqrt{q^2+\b^2r^{2(n-1)}}\right)^2+{\cal O}\left(\frac{1}{\e}\right).
\end{equation}
For flat horizon solution ($\ve=0$) we obtain:
\begin{equation}
UW\simeq\e^2\left(\frac{\L-2\b^2}{2\al}+\frac{\b\e}{\al}r^{1-n}\sqrt{q^2+\b^2r^{2(n-1)}}\right)^2+{\cal O}(\e).
\end{equation}

\begin{figure}
\centerline{\includegraphics[scale=0.33,clip]{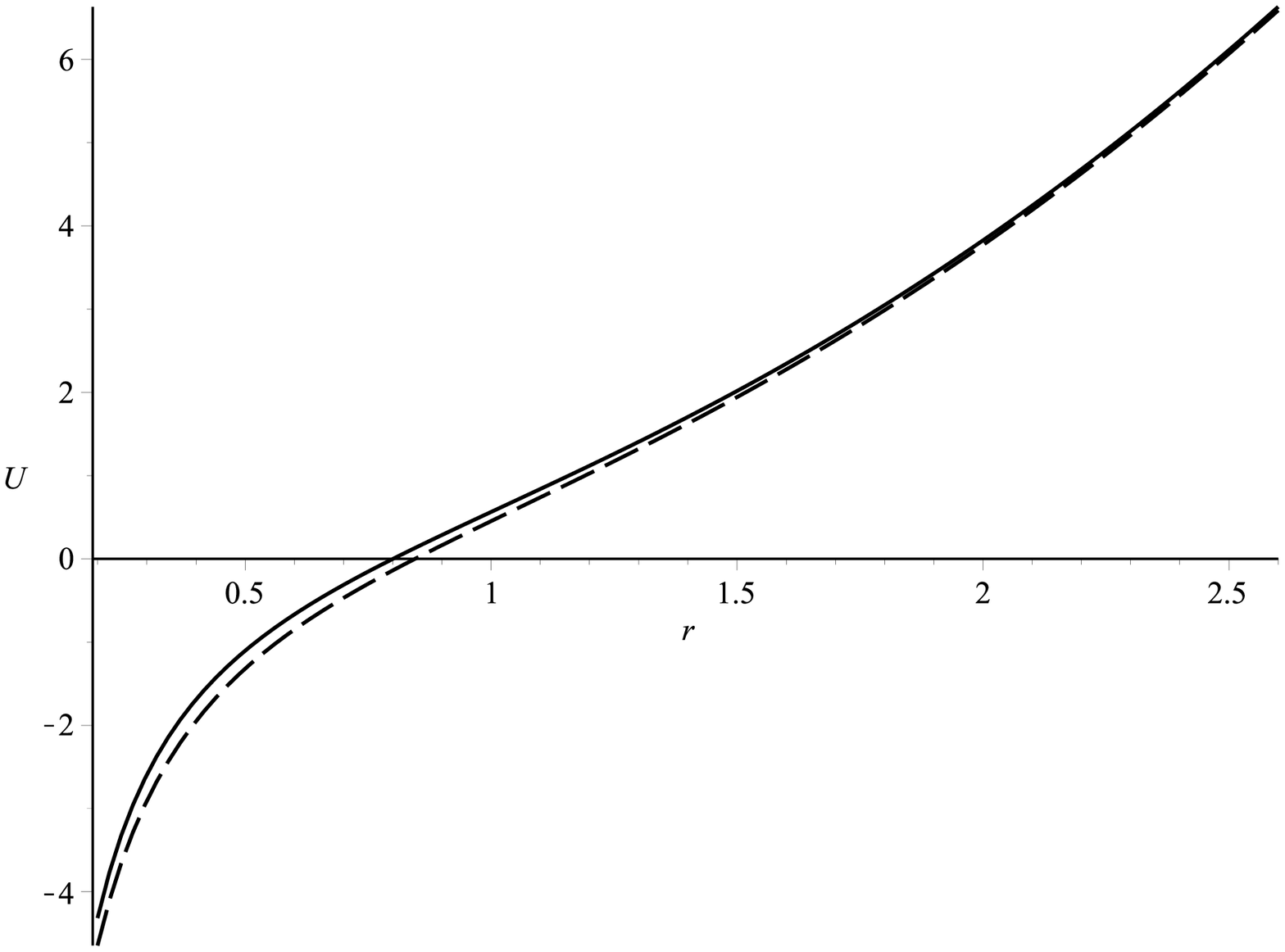}\includegraphics[scale=0.33,clip]{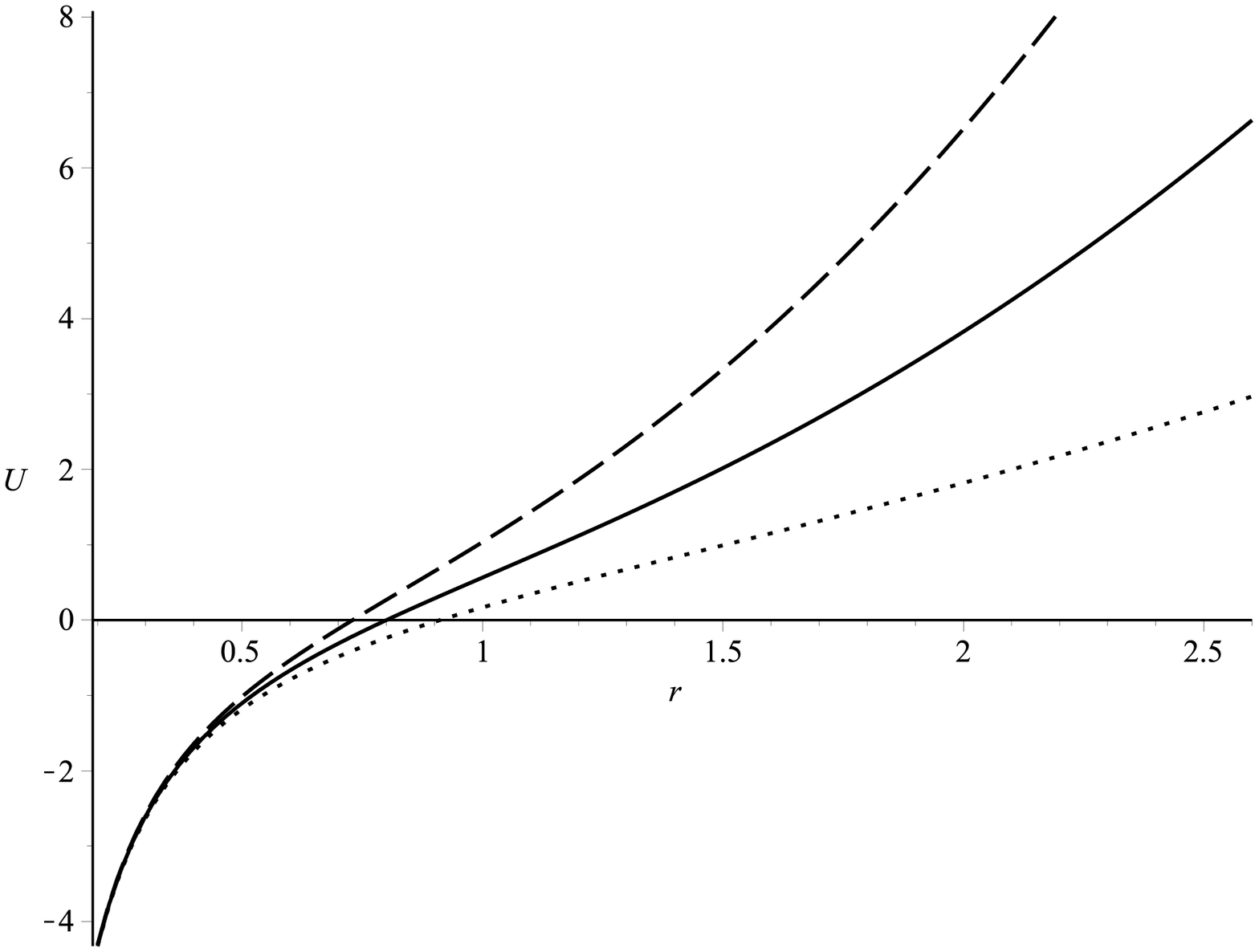}}
\caption{Metric functions $U(r)$ for different values of parameter $\beta$ (the left graph) and different values of the cosmological constant $\L$ (the right one). For all the graphs we have $n=3$, $\ve=1$, $\al=0.2$, $\eta=0.4$, $\mu=1$, $q=0.2$. For the left graph the other parameters are equal to $\L=-2$, $\beta=1$ (solid curve) $\beta=10$ (dashed curve). For the right graph $\beta=1$, and $\L=-1$ (dotted curve), $\L=-2$ (solid curve) and $\L=-3$ (dashed curve).}\label{metr_f_graph}
\end{figure}

We have analyzed the behaviour of the metric function $U(r)$  for different distances $r$ and various values of some parameters. Here we also give graphical representation which might make the given above analysis more transparent. The figure [\ref{metr_f_graph}] shows that the change of the parameter $\beta$ (parameter of nonlinearity of the gauge field) affects weakly on the behaviour of the metric function (at least this effect is small in the domain of change of the parameters we have used here). The influence of the cosmological constant $\L$ as it follows even from the above analysis is more important for large distances, where the AdS-term becomes dominant, but for small distances the influence of the AdS-term is negligibly small.

Some particular interest might be also in considering of the regime of small $\beta$, namely for the metric function $U(r)$ we have:
\begin{equation}\label{metr_small_b}
U(r)\simeq\ve-\frac{\mu}{r^{n-2}}-\frac{2\L}{n(n-1)}r^2-\frac{2\b(\al-\L\e)}{\al(n-1)}\frac{q}{r^{n-3}}-\frac{2\b(\al+\L\e)}{\al(n-1)}\frac{qd}{r^{n-2}}\arctan{\left(\frac{r}{d}\right)}+{\cal{O}}(\b^2),
\end{equation}
and for the product of the metric functions we write:
\begin{equation}
UW\simeq\frac{\left((\al-\L\e)r^2+\ve\e(n-1)(n-2)\right)}{(2\al r^2+\ve\e(n-1)(n-2))^2}\left((\al-\L\e)r^2+\ve\e(n-1)(n-2)-4\e q r^{3-n}\right)+{\cal{O}}(\b^2). 
\end{equation}
We point out that for the flat horizon case ($\ve=0$) the last term in the  asymptotic relation (\ref{metr_small_b}) does not appear. It should be also noted that given above two relations are valid for small and intermediate distances, since the product $\b r^{2(n-1)}$ which is present in general relation for the metric functions might become large for corresponding large values of $r$. 

To obtain information about coordinate and physical singularities of the metric Kretschmann scalar at different points should be examined. In general it takes the form:
\begin{equation}\label{Kr_scalar}
R_{\mu\nu\k\l}R^{\mu\nu\k\l}=\frac{1}{UW}\left(\frac{d}{dr}\left[\frac{U'}{\sqrt{UW}}\right]\right)^2+\frac{(n-1)}{r^2W^2}\left(\frac{(U')^2}{U^2}+\frac{(W')^2}{W^2}\right)+\frac{2(n-1)(n-2)}{r^4W^2}(\ve W-1)^2.
\end{equation}
One can verify easily that at the horizon point $r_+$, namely when $U(r_+)=0$ the Kretschmann scalar (\ref{Kr_scalar}) is nonsingular, it means that the horizon points are the points with a coordinate singularity as it should be for a black hole. To investigate the behavior of the metric at the origin and at the infinity one should use corresponding asymptotic relations that has been obtained previously. At large distances when $r\rightarrow\infty$ the Kretschmann scalar is as follows:
\begin{equation}
R_{\mu\nu\k\l}R^{\mu\nu\k\l}\simeq \frac{8(n+1)\al^2}{n(n-1)^2\e^2}
\end{equation}
It should be noted that the relation for Kretschmann scalar at the infinity is completely the same as it is for a chargeless solution \cite{Stetsko_slr} or with nonlinear electromagnetic field \cite{Stetsko_PRD19}. This similarity is caused by the same asymptotic behaviour of all the metrics at the infinity.  Since the metric functions have different behaviour in the domain close to the origin of coordinates for diverse dimensions of space $n$ and values of $\ve$ it means that the Kretschmann scalar (\ref{Kr_scalar}) should be examined separately for all these cases. Firstly, we consider the situation when $n\geqslant 5$ and $\ve=\pm 1$. Having substituted the leading terms of the metric functions given by the relations (\ref{u_r0_c}) and (\ref{UW_r0_c}) into the relation (\ref{Kr_scalar}) and after little algebra we obtain: 
\begin{equation}
R_{\mu\nu\k\l}R^{\mu\nu\k\l}\simeq \frac{4(n-1)^2(n-2)(n-3)^2}{(n-4)r^4}.
\end{equation}  
It should be pointed out that the obtained relation does not contain the integration constant $q$ nor the parameter $\b$ which are present in the leading terms of the metric functions. We also remark here that for power-law field as well as for the linear one \cite{Stetsko_PRD19} the Kretschmann scalar at the origin has similar dependence $\sim 1/r^4$ and does not depend on the charge parameter $q$.  Now we derive the Kretschmann scalar when $n=4$ and $\ve\neq 0$. Here it is necessary to utilize the relations (\ref{u_r0_c4}) and (\ref{UW_r0_c}). As a result we arrive at the expression:
\begin{equation}\label{Kr_sc_4_c}
R_{\mu\nu\k\l}R^{\mu\nu\k\l}\simeq\frac{40}{r^4}\ln^2\left(1+\frac{d^2}{r^2}\right).
\end{equation}
We note here that the written above expression does not depend on the parameters $q$ and $\b$ analogously as it was in the previous case ($n\geqslant 5$), but it has a bit stronger singular behaviour at the origin due to the presence of a divergent logarithmic factor. In three dimensional space ($n=3$) we have to use the the relation (\ref{u_r0_3}) and the corresponding relation for the function $W$ and as a result the Kretschmann scalar in the vicinity of the origin takes the form:
\begin{equation}\label{Kr_sc_n3}
R_{\mu\nu\k\l}R^{\mu\nu\k\l}\simeq\frac{\mu^2}{\left(1-\frac{\b q}{\ve}\right)^4 r^6}
\end{equation}  
We conclude that in three dimensional space the singularity of the Kretschmann scalar is the strongest and it is caused by the other term than in the previously analyzed cases when $n\geqslant 4$. When $q=0$ the relation (\ref{Kr_sc_n3}) is completely the same as it was for the neutral case \cite{Stetsko_slr}.

Now we consider flat horizon geometry ($\ve=0$) and we should use the asymptotic relation (\ref{u_r0_f}) and corresponding relation for the metric function $W$. We remark that for the flat horizon geometry we do not consider the cases of various dimensions separately because the behaviour of the metric functions is defined by similar relation for all the dimensions. As a result we obtain:
\begin{equation}\label{Kr_sc_r_f}
R_{\mu\nu\k\l}R^{\mu\nu\k\l}\simeq\frac{2(n-1)(n-2)}{r^4}
\end{equation}
and here similarly to the investigated above cases of nonflat geometry  the behaviour of the Kretschmann scalar does not depend on the parameters $q$ and $\b$, but in contrast with the nonflat cases the relation (\ref{Kr_sc_r_f}) is valid for all the dimensions $n\geqslant 3$.

We also note here that for the given solutions there is an additional point where the Kretschmann scalar is singular, namely this is the point where $W(r)=0$. It is easy to check that such a point exists, it can be verified by virtue of the relation (\ref{UW_prod}), namely when the enumerator of that relation is equal to zero. This singularity point takes place in the interior region, since the condition we impose when we set to zero the enumerator of the relation (\ref{UW_prod}) is compatible with the condition we have imposed on the square of the derivative of the scalar field $(\vp')^2$ (\ref{fi_der_2}). It should be noted, that at this point the components of the metric tensor are not singular, but the singularity of the Kretschmann scalar appears due to the fact that the determinant of the metric tensor is equal to zero here.
\subsection{Electromagnetic field potential}
The important feature of a charged black hole is its gauge potential, namely it has key role when one tries to derive the first law of black holes thermodynamics. The evident form of the gauge potential can be found easily when one uses the relation for the gauge field (\ref{EM_field}). So, we can write:
\begin{equation}\label{gauge_pot}
A_0(r)\equiv\psi=\psi_0+\int\frac{q\b}{\sqrt{q^2+\b^2r^{2(n-1)}}}\sqrt{UW}dr.
\end{equation}
Taking into account the relation for the product of the metric functions (\ref{UW_prod}) one can perform the integration and write the gauge potential in the form:
\begin{eqnarray}\label{gauge_pot_odd}
\nonumber A_0(r)\equiv\psi(r)=\psi_0-\frac{q}{2(n-2)\al}\frac{(\al-\L\e+2\b^2\e)}{r^{n-2}}{_{2}F_{1}}\left(\frac{1}{2},\frac{n-2}{2(n-1)};\frac{3n-4}{2(n-1)};-\frac{q^2}{\b^2}r^{2(1-n)}\right)-\\\nonumber\frac{qd^2}{2\al}\frac{(\al+\L\e-2\b^2\e)}{r^n}\sum^{+\infty}_{j=0}\frac{(-1)^j}{n+2j}\left(\frac{d}{r}\right)^{2j}{_{2}F_{1}}\left(\frac{1}{2},\frac{n+2j}{2(n-1)};\frac{3n+2(j-1)}{2(n-1)};-\frac{q^2}{\b^2}r^{2(1-n)}\right)-\\\frac{\b^2\e q}{\al}\left(\sum^{(n-5)/2}_{j=0}\frac{(-1)^jr^{4-n+2j}}{(4-n+2j)d^{2(j+1)}}+\frac{(-1)^{\frac{n-3}{2}}}{d^{n-2}}\arctan\left(\frac{r}{d}\right)\right)
\end{eqnarray}
for odd $n$ and large $r$ and
\begin{eqnarray}\label{gauge_pot_even}
\nonumber A_0(r)\equiv\psi(r)=\psi_0-\frac{q}{2(n-2)\al}\frac{(\al-\L\e+2\b^2\e)}{r^{n-2}}{_{2}F_{1}}\left(\frac{1}{2},\frac{n-2}{2(n-1)};\frac{3n-4}{2(n-1)};-\frac{q^2}{\b^2}r^{2(1-n)}\right)-\\\nonumber\frac{qd^2}{2\al}\frac{(\al+\L\e-2\b^2\e)}{r^n}\sum^{+\infty}_{j=0}\frac{(-1)^j}{n+2j}\left(\frac{d}{r}\right)^{2j}{_{2}F_{1}}\left(\frac{1}{2},\frac{n+2j}{2(n-1)};\frac{3n+2(j-1)}{2(n-1)};-\frac{q^2}{\b^2}r^{2(1-n)}\right)-\\\frac{\b^2\e q}{\al}\left(\sum^{(n-6)/2}_{j=0}\frac{(-1)^jr^{4-n+2j}}{(4-n+2j)d^{2(j+1)}}+\frac{(-1)^{\frac{n-2}{2}}}{2d^{n-2}}\ln\left(1+\frac{d^2}{r^2}\right)\right)
\end{eqnarray}
for even $n$ and large $r$ respectively. We note that almost all the terms which depend on the radial coordinate $r$ go to zero at infinity and the only exclusion is the $\arctan(r/d)$ term in the case of odd $n$ but this term is finite at the infinity, so the value of the gauge potential at the infinity is mainly defined by the constant of integration $\psi_0$ that we have introduces in our relation. This constant can be taken arbitrary, but one can fix it imposing a condition on the potential in arbitrary point.  We note that for small $r$ another representation for hypergeometric function should be used. Namely, for odd $n$ we can write:
 \begin{eqnarray}\label{g_pot_odd_sm}
\nonumber A_0(r)\equiv\psi(r)=\psi_0+\frac{\b(\al-\L\e+2\b^2\e)}{2\al}r{_{2}F_{1}}\left(\frac{1}{2},\frac{1}{2(n-1)};\frac{2n-1}{2(n-1)};-\frac{\b^2}{q^2}r^{2(n-1)}\right)+\\\nonumber\frac{\b(\al+\L\e-2\b^2\e)}{2\al}r\sum^{+\infty}_{j=0}\frac{(-1)^j}{2j+1}\left(\frac{r}{d}\right)^{2j}{_{2}F_{1}}\left(\frac{1}{2},\frac{2j+1}{2(n-1)};\frac{2(j+n)-1}{2(n-1)};-\frac{\b^2}{q^2}r^{2(n-1)}\right)-\\\frac{\b^2\e q}{\al}\left(\sum^{(n-5)/2}_{j=0}\frac{(-1)^jr^{4-n+2j}}{(4-n+2j)d^{2(j+1)}}+\frac{(-1)^{\frac{n-3}{2}}}{d^{n-2}}\arctan\left(\frac{r}{d}\right)\right)
\end{eqnarray}
and for even $n$ we obtain:
 \begin{eqnarray}\label{g_pot_even_sm}
\nonumber A_0(r)\equiv\psi(r)=\psi_0+\frac{\b(\al-\L\e+2\b^2\e)}{2\al}r{_{2}F_{1}}\left(\frac{1}{2},\frac{1}{2(n-1)};\frac{2n-1}{2(n-1)};-\frac{\b^2}{q^2}r^{2(n-1)}\right)+\\\nonumber\frac{\b(\al+\L\e-2\b^2\e)}{2\al}r\sum^{+\infty}_{j=0}\frac{(-1)^j}{2j+1}\left(\frac{r}{d}\right)^{2j}{_{2}F_{1}}\left(\frac{1}{2},\frac{2j+1}{2(n-1)};\frac{2(j+n)-1}{2(n-1)};-\frac{\b^2}{q^2}r^{2(n-1)}\right)-\\\frac{\b^2\e q}{\al}\left(\sum^{(n-6)/2}_{j=0}\frac{(-1)^jr^{4-n+2j}}{(4-n+2j)d^{2(j+1)}}+\frac{(-1)^{\frac{n-2}{2}}}{2d^{n-2}}\ln\left(1+\frac{d^2}{r^2}\right)\right)
\end{eqnarray}
The written above relations show that near the origin of coordinates the gauge potential $A_0(r)$ is singular when $n\geqslant 4$, namely when $n=4$ the potential has a logarithmic divergence at the origin whereas for $n>4$ we have power-law singularity. It should be pointed out that even the gauge field (\ref{EM_field}) is singular at the origin if $n\geqslant 4$, the only case when the gauge field and the potential are not singular is $n=3$. We also emphasize that the behaviour of the gauge field and potential in our case is completely different from the situation which takes place in standard General Relativity with a gauge field described by a Born-Infeld type of action, where the gauge field and potential have nonsingular behaviour for all values of $n$.

As we have noted above if we impose some condition on the potential it allows us to fix the integration constant $\psi_0$. We impose that at the horizon $r_+$ the potential $A_0(r)$ is equal to zero ($A(r_+)=0$). This requirement is not necessary, but it will be convenient for us when we use the Wald procedure in the following section. Since we have imposed that $A(r_+)=0$ this condition allows us to represent the constant $\psi_0$ as a function of the horizon radius $r_+$ and the other parameters such as $q$, $\L$, $\al$, $\b$, $\e$ and $n$, namely for the odd $n$ we can write:
\begin{eqnarray}\label{psi_0_odd}
\nonumber\psi_0=-\frac{\b(\al-\L\e+2\b^2\e)}{2\al}r_+{_{2}F_{1}}\left(\frac{1}{2},\frac{1}{2(n-1)};\frac{2n-1}{2(n-1)};-\frac{\b^2}{q^2}r^{2(n-1)}_+\right)-\\\nonumber\frac{\b(\al+\L\e-2\b^2\e)}{2\al}r_+\sum^{+\infty}_{j=0}\frac{(-1)^j}{2j+1}\left(\frac{r_+}{d}\right)^{2j}{_{2}F_{1}}\left(\frac{1}{2},\frac{2j+1}{2(n-1)};\frac{2(j+n)-1}{2(n-1)};-\frac{\b^2}{q^2}r^{2(n-1)}_+\right)+\\\frac{\b^2\e q}{\al}\left(\sum^{(n-5)/2}_{j=0}\frac{(-1)^jr^{4-n+2j}_+}{(4-n+2j)d^{2(j+1)}}+\frac{(-1)^{\frac{n-3}{2}}}{d^{n-2}}\arctan\left(\frac{r_+}{d}\right)\right).
\end{eqnarray}
We point out that here we have used the relation (\ref{g_pot_odd_sm}), which gives the evident form of the gauge potential for relatively small distances, where the horizon $r_+$ is located. For the case of even $n$ instead of (\ref{g_pot_odd_sm}) the relation (\ref{g_pot_even_sm}) should be utilized and as a result we can write:
\begin{eqnarray}\label{psi_0_even}
\nonumber\psi_0=-\frac{\b(\al-\L\e+2\b^2\e)}{2\al}r_+{_{2}F_{1}}\left(\frac{1}{2},\frac{1}{2(n-1)};\frac{2n-1}{2(n-1)};-\frac{\b^2}{q^2}r^{2(n-1)}_+\right)-\\\nonumber\frac{\b(\al+\L\e-2\b^2\e)}{2\al}r_+\sum^{+\infty}_{j=0}\frac{(-1)^j}{2j+1}\left(\frac{r_+}{d}\right)^{2j}{_{2}F_{1}}\left(\frac{1}{2},\frac{2j+1}{2(n-1)};\frac{2(j+n)-1}{2(n-1)};-\frac{\b^2}{q^2}r^{2(n-1)}_+\right)+\\\frac{\b^2\e q}{\al}\left(\sum^{(n-6)/2}_{j=0}\frac{(-1)^jr^{4-n+2j}_+}{(4-n+2j)d^{2(j+1)}}+\frac{(-1)^{\frac{n-2}{2}}}{2d^{n-2}}\ln\left(1+\frac{d^2}{r^2_+}\right)\right)
\end{eqnarray}

If $\ve=0$ the gauge potential takes a simpler form, namely for large $n$ we can write:
\begin{equation}\label{g_pt_e0_lar}
A_0(r)\equiv\psi=\psi_0+\frac{\b^2\e q}{(n-2)\al}r^{2-n}-\frac{(\al-\L\e+2\b^2\e)q}{2(n-2)\al r^{n-2}}{_{2}F_{1}}\left(\frac{1}{2},\frac{2-n}{2(1-n)};\frac{4-3n}{2(1-n)};-\frac{q^2}{\b^2}r^{2(1-n)}\right)
\end{equation}
and for small $r$ we obtain:
\begin{equation}\label{g_pt_e0_sm}
A_0(r)\equiv\psi=\psi_0+\frac{\b^2\e q}{(n-2)\al}r^{2-n}+\frac{\b(\al-\L\e+2\b^2\e)}{2\al}r{_{2}F_{1}}\left(\frac{1}{2},\frac{1}{2(n-1)};\frac{2n-1}{2(n-1)};-\frac{\b^2}{q^2}r^{2(n-1)}\right).
\end{equation}
It follows from the latter two relations that the gauge potential has power-law singularity at the origin for all $n\geqslant3$ and at the infinity it equals to the constant $\psi_0$.

Having the explicit form of the gauge field (\ref{EM_field}) and using the Gauss law we calculate total charge of the black hole which is of the crucial importance for black hole thermodynamics. The Gauss law for the Born-Infeld electrodynamics takes the following form:
\begin{equation}
Q=\frac{1}{4\pi}\int_{\Sigma}\left(1+\frac{F_{\k\l}F^{\k\l}}{2\b}\right)^{-\frac{1}{2}}*F
\end{equation}
and here $*F$ denotes the Hodge dual of electromagnetic field form $F$ and the integral is taken over a closed $n-1$--dimensional hypersurface $\Sigma$. After calculation of the latter integral one arrives at:
\begin{equation}
Q=\frac{\omega_{n-1}}{4\pi}q,
\end{equation}
where $\omega_{n-1}$ is the hypersurface area of a ``unit'' hypersurface of constant curvature (it would be surface area of a unit hypersphere in case of spherical symmetry). For nonspherical geometry of horizon it is convenient to define the total electric charge per unit area which can be written in the form:
\begin{equation}
\bar{Q}=\frac{1}{\omega_{n-1}}Q.
\end{equation}
The electric potential measured at the infinity with respect to the horizon can be represented in the form:
\begin{equation}
\Phi_q=A_{\mu}\chi^{\mu}\Big|_{+\infty}-A_{\mu}\chi^{\mu}\Big|_{r_+},
\end{equation}
where $\chi^{\mu}$ is a timelike Killing vector null on the event horizon   and we have taken it to be the time translation vector $\chi^{\mu}=\partial/\partial t$. After calculation we arrive at the relation:
\begin{equation}
\Phi_q=\psi_0,
\end{equation}
where now the parameter $\psi_0$ is determined as a function of the horizon radius $r_+$ and the other parameters due to the relations (\ref{psi_0_odd}) and (\ref{psi_0_even}) for odd and even $n$ respectively. It should be pointed out that one can impose that $A_{\mu}$ is zero at the infinity, but it would give rise to the same expression for the measured electric potential $\Phi_q$.

\section{Black hole thermodynamics}
In this section we derive and investigate main thermodynamic relations of the black hole solutions obtained in the previous section. One of the most important thermodynamic quantities of the black hole is its temperature which can be obtained in the same way as it is done in General Relativity, namely its definition is based on the notion of surface gravity which can represented in the following form:
\begin{equation}\label{surf_grav}
\kappa^2=-\frac{1}{2}\nabla_{a}{\chi}_b\nabla^{a}{\chi}^{b},
\end{equation}
and here $\bar{\chi}_a$ is a Killing vector, which should be null on the event horizon. Similarly as in the previous section we take the vector of time translation ${\chi}^a=\partial/\partial t$. One can calculate the surface gravity (\ref{surf_grav}) and substituting it in the definition of the temperature one arrives at the relation:
\begin{equation}\label{BH_temp}
T=\frac{\k}{2\pi}=\frac{1}{4\pi}\frac{U\rq{}(r_+)}{\sqrt{U(r_+)W(r_+)}}
\end{equation}  
where $r_+$ is the radius of the event horizon of the black hole. It should be pointed out that nonetheless on the complicated structure of the metric function $U(r)$ the evident form of which contains hypergeometric functions, the temprature can be represented in a relatively compact form which comprises just of rational and irrational functions of the horizon radius $r_+$. After all the calculations the temperature can be represented in the following form:
\begin{eqnarray}\label{temp_BI_gen} 
\nonumber T=\frac{1}{4\pi}\frac{2\al r^2_{+}+\ve\e(n-1)(n-2)}{(\al-\L\e+2\b^2\e)r^2_{+}+\ve\e(n-1)(n-2)-2\b\e r^{3-n}_{+}\sqrt{q^2+\b^2r^{2(n-1)}_{+}}}\times\\\nonumber\left[\frac{(n-2)\ve}{r_+}-\frac{2\b(\al-\L\e+2\b^2\e)}{\al(n-1)r^{n-2}_+}\sqrt{q^2+\b^2r^{2(n-1)}_+}+\frac{(\al+\L\e-2\b^2\e)^2}{2(n-1)\al\e}\frac{r^3_+}{r^2_{+}+d^2}-\right.\\\left.\frac{2(\L-2\b^2)}{n-1}r_{+}-\frac{2\b(\al+\L\e-2\b^2\e)d^2}{\al(n-1)r^{n-2}_+}\frac{\sqrt{q^2+\b^2r^{2(n-1)}_+}}{r^2_{+}+d^2}+\frac{2\b^2\e\left(q^2+\b^2r^{2(n-1)}_+\right)}{\al(n-1)r^{2n-5}_{+}(r^2_{+}+d^2)}\right].
\end{eqnarray}
In the limit when $\b\rightarrow\infty$ we recover the relation for the temperature for linear Maxwell field, namely we arrive at \cite{Stetsko_PRD19}:
\begin{eqnarray}\label{temp_lin_f}
\nonumber T=\frac{1}{4\pi}\frac{2\al r^2_{+}+\ve\e(n-1)(n-2)}{(\al-\L\e)r^2_{+}+\ve\e(n-1)(n-2)-\e q^2r^{2(2-n)}_{+}}\left[\frac{(n-2)\ve}{r_+}+\frac{(\al-\L\e)^2}{2\al\e}r_{+}-\frac{2q^2}{(n-1)r^{2n-3}_{+}}\right.\\\left.+\frac{r_+}{\al(n-1)(r^2_{+}+d^2)}\left((\al+\L\e)q^2r^{2(2-n)}_{+}-\frac{(\al+\L\e)^2}{2\e}d^2+\frac{\e q^4}{2}r^{2(3-2n)}_{+}\right)\right].
\end{eqnarray}
Here we remark, that the relation for the temperature for linear Maxwell field given in our previous work \cite{Stetsko_PRD19} was written in a bit different form, but some simple transformations allows us to obtain the relation (\ref{temp_lin_f}).

The obtained relations for the temperature (\ref{temp_BI_gen}) for arbitrary $\b$, as well as its particular case (\ref{temp_lin_f}) for linear field are not so simple to comprehend their behaviour in full details, but nevertheless some general conclusions can be made just looking at the given above relations. First of all for large horizon radii $r_+$ the dominant term in both cases is linear over $r_+$, namely we arrive at the relation $T\simeq (\al-\L\e)r_+/4\pi\e$ for arbitrary $\b$. The fact that the leading term of the asymptotic relation does not depend on the parameter $\b$ can be explained by the domination of the AdS-term in this case and also thanks to the circumstance that Born-Infeld modification of gauge action was introduced to eliminate divergent behaviour of the gauge field for small distances, whereas for the large ones the electromagnetic field behaves almost in the same way as for the linear field case. For small horizon radius the behaviour of the temperature is different for arbitrary finite $\b$ and for the limit case of linear field ($\b\rightarrow\infty$), namely for the first case we have $T\sim\b r^{2-n}_+$ whereas for the latter one it behaves as $T\sim r^{3-2n}_+$ so we conclude that for linear field the dependence of the temperature for small radius of horizon is stronger for linear field, but this conclusion is expectable since as we have already mentioned Born-Infeld theory was introduced to modify the electromagnetic field in a way to make it finite at the origin, thus it does not takes place here in the theory with nonminimal derivative coupling apart of the case when $n=3$. We also pay attention to the case $n=3$ here, namely when $\b\neq 0$ and $\ve\neq 0$ for small horizon radius the temperature might change its sign depending on the relation between $\b$ and $q$.
     
\begin{figure}
\centerline{\includegraphics[scale=0.33,clip]{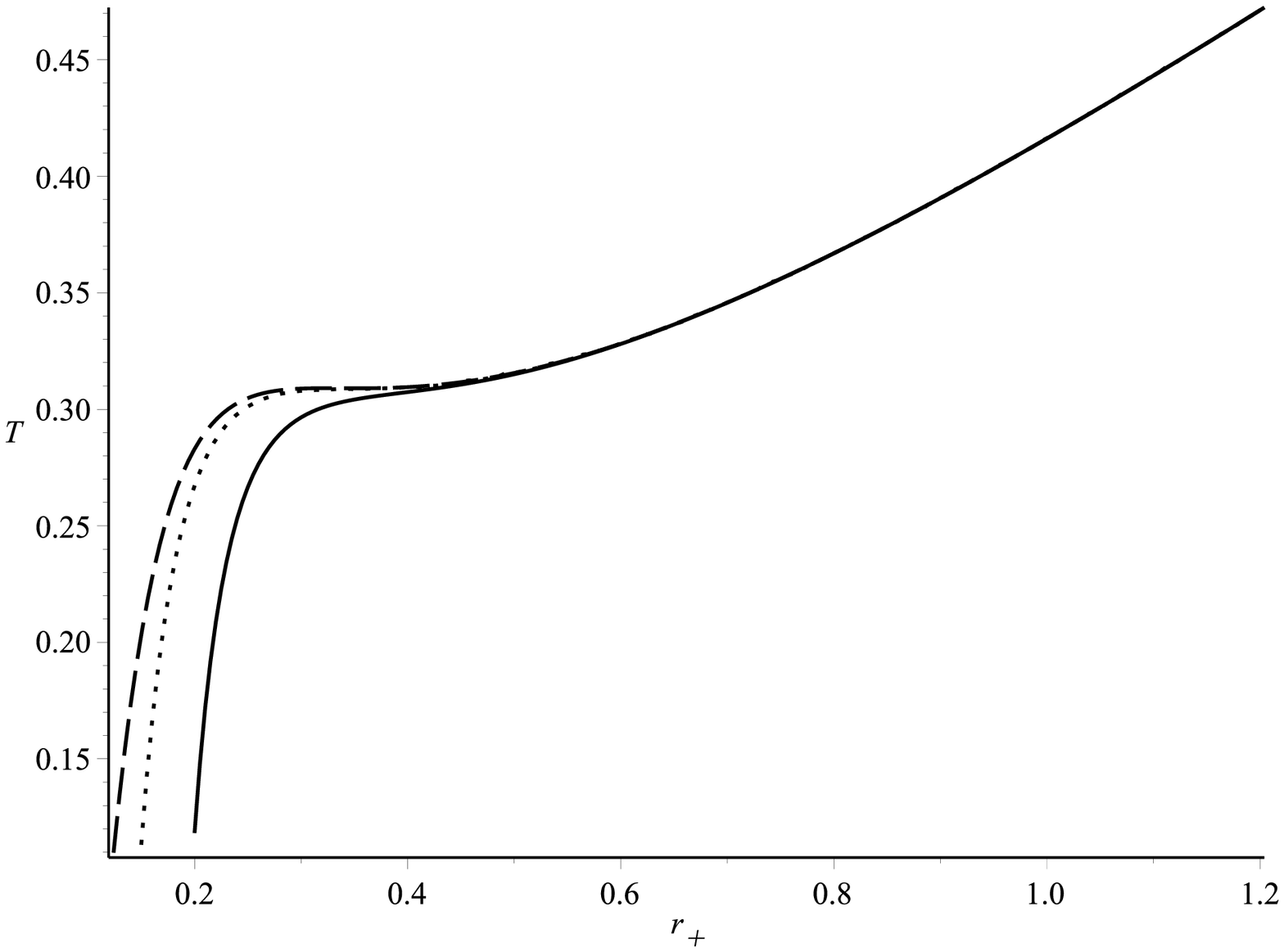}\includegraphics[scale=0.33,clip]{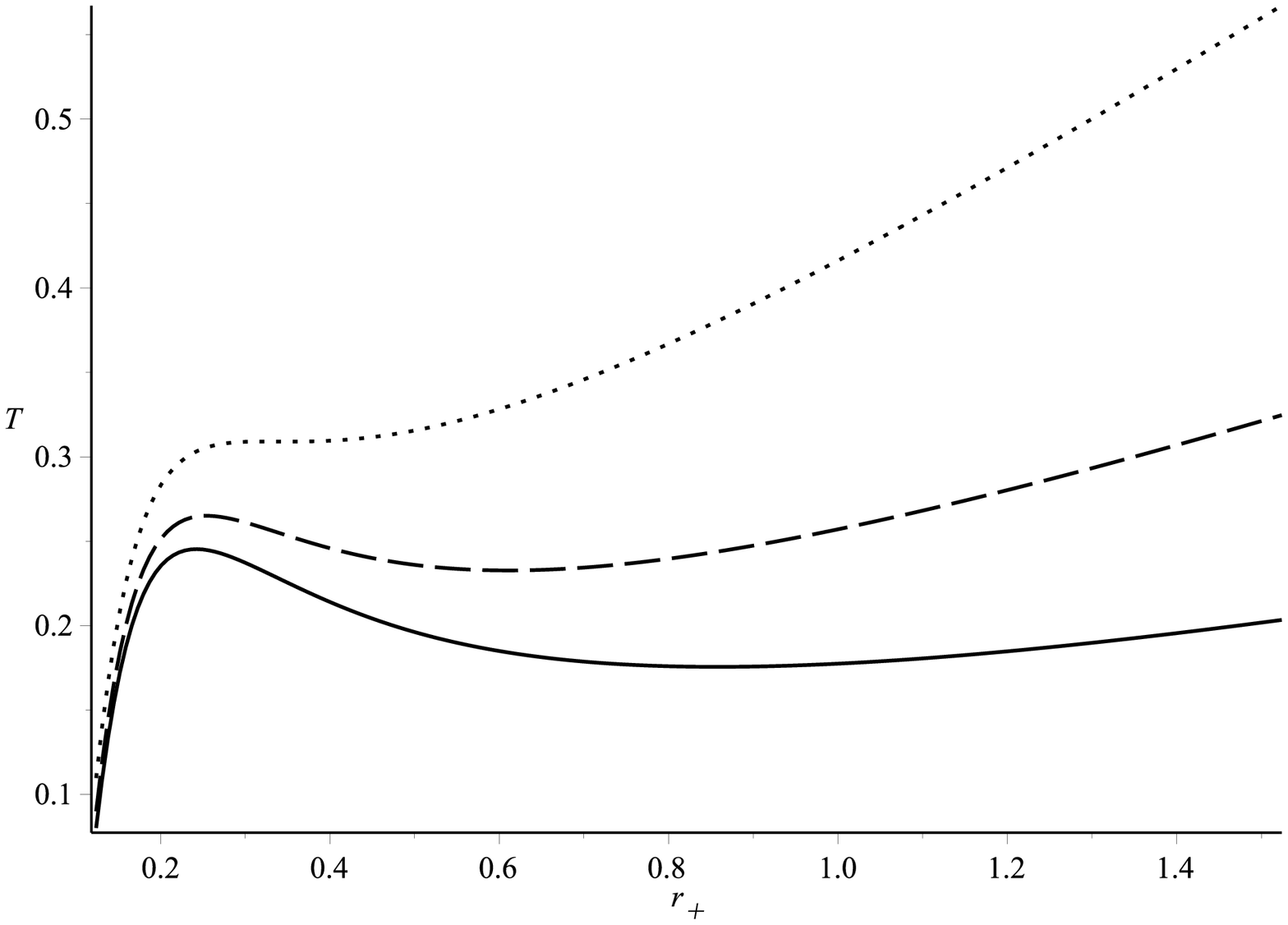}}
\caption{Temperature $T$ as a function of the horizon radius $r_+$ for different values of parameter $\beta$ (the left graph) and different values of the cosmological constant $\L$ (the right one). For all the graphs we have $n=3$, $\ve=1$, $\al=0.2$, $\eta=0.4$, $q=0.2$. For the left graph the other parameters are equal to $\L=-8$, $\beta=8$ (dashed line), $\beta=50$ (dotted line) and solid line represents linear field case ($\b\rightarrow\infty$). For the right graph $\beta=8$, and $\L=-2$ (solid line), $\L=-4$ (dashed line) and $\L=-8$ (dotted line).}\label{temp_graph}
\end{figure}

We have analyzed the dependences for the temperatures (\ref{temp_BI_gen}) and (\ref{temp_lin_f}) for some particular cases of $r_+$, namely for its extremely large and small values, to understand  behaviour of the temperatures better for some intermediate values of $r_+$ we demonstrate it graphically. Figure [\ref{temp_graph}] shows this dependence for various values of $\b$, when the other parameters are held fixed (the left graph) and for various values of the cosmological constant $\L$ (the right graph). We can conclude that the variation of the parameter $\b$ affects considerably on the temperature for small horizon radius, whereas for large values of $r_+$ this influence is negligibly small and the behaviour is mainly defined by the AdS-terms in all the cases. Variation of the parameter $\L$ has substantial influence on the temperature for large $r_+$, whereas for small $r_+$ its contribution becomes negligibly small. The function $T(r_+)$ might be nonmonotonous, what is better seen on the right graph, this fact might give us some critical behaviour in extended thermodynamic phase space  similarly as it is done for charged black holes in the framework of standard General Relativity \cite{Kubiznak_JHEP12,Gunasekaran_JHEP12}, but this issue will be investigated elsewhere.  
\subsection{Wald's approach and entropy of black hole}
There are several approaches to define entropy of a black hole, some of them take their roots in earlier work of Gibbons and Hawking \cite{Gibbons_PRD77}, other approaches are based on Wald's procedure (or method) which can be treated as a generalization of Noether method to derive conserved quantities in gravity \cite{Wald_PRD93,Iyer_PRD94}. It should be pointed out that Wald's procedure applicable to quite general diffeomorphism-invariant theories. It has been applied to numerous black hole solutions in the framework of standard General Relativity as well as its generalizations \cite{Feng_JHEP15,Feng_PRD16,Stetsko_PRD19,Liu_PLB14,Lu_JHEP15,Fan_JHEP15}.  In Horndeski Gravity the Wald's procedure appears to be a consistent approach to obtain the first law of black hole thermodynamics. We point out here that the scalar potential in Horndeski Gravity has singular behaviour, but the Wald's procedure takes this fact into account \cite{Feng_JHEP15,Feng_PRD16}. We also note that Wald's procedure allowed to derive reasonable and in some sense universal relations for back hole's entropy \cite{Feng_JHEP15,Feng_PRD16,Stetsko_PRD19}.
 
Here we also utilize Wald's procedure to derive a relation for black hole's entropy and firstly we describe the keypoints of this approach. Assuming that we have a Lagrangian ${\cal L}$ of a system, we can perform its variation and as a result we write:
\begin{equation}
\delta {\cal L}=e.o.m.+\sqrt{-g}\nabla_{\mu}J^{\mu},
\end{equation}
where $e.o.m.$ represents the terms which give equations of motion for the system and the term $\sqrt{-g}\nabla_{\mu}J^{\mu}$ gives rise to the so called surface term in the action integral because of full divergence of the last term, the $J^{\mu}$ denotes the surface ``current''. Using the obtained relation for the current $J^{\mu}$ one can construct ``current'' one-form: $J_{(1)}=J_{\nu}dx^{\nu}$ and its Hodge dual: $\Theta_{(n)}=*J_{(1)}$. Having utilized infinitesimal diffeomorphism given by the vector $\delta x^{\mu}=\xi^{\mu}$ one can define the form:
\begin{equation}\label{diff_forms}
J_{(n)}=\Theta_{(n)}-i_{\xi}*{\cal L}=e.o.m.-d*J_{(2)},
\end{equation}
and here $i_{\xi}*{\cal L}$ denotes contraction of the vector field $\xi^{\mu}$ with the dual of the form $*{\cal L}$. In case the equations of motion are fulfilled (on-shell condition) we can conclude that the form $J_{(n)}$ is exact, namely $J_{(n)}=dQ_{(n-1)}$, where $Q_{(n-1)}=-*J_{(2)}$. The written above relation (\ref{diff_forms}) allows to obtain thermodynamic relations, namely the first law of black hole thermodynamics if the infinitesimal diffeomorphism vector $\xi^{\mu}$ is taken to be a Killing vector null at the horizon. It was shown by Wald that the variation of gravitational Hamiltonian can be represented in the form:
\begin{equation}\label{var_hamilt}
\delta{\cal H}=\frac{1}{16\pi}\left(\delta \int_{c}J_{(n)}-\int_{c}d(i_{\xi}\Theta_{(n)})\right)=\frac{1}{16\pi}\int_{\Sigma_{(n-1)}}(\delta Q-i_{\xi}\Theta_{(n)}),
\end{equation}
where $c$ denotes $n$--dimensional Cauchy surface and $\Sigma_{(n-1)}$ is its $n-1$--dimensional boundary which consist of two parts: one on the event horizon and the other at the infinity. The first law of black hole thermodynamics can be obtained from the relation:
\begin{equation}\label{variat_inf_hor}
\delta{\cal H}_{\infty}=\delta{\cal H}_{+},
\end{equation}
we note that the in the latter relation the variation in the left hand side is taken at the infinity and in the right hand side is taken at the events horizon.

Using written above relations we can calculate the variation of Hamiltonian and consequently derive the first law. For minimally  coupled part of the action with Born-Infeld term we can write:
\begin{equation}\label{var_min}
(\delta Q-i_{\xi}\Theta)_{min}=r^{n-1}\sqrt{UW}\left(\frac{(n-1)}{rW^2}\delta W+\frac{2}{UW}\frac{1}{\left(1-\frac{(\psi')^2}{\beta^2UW}\right)^{\frac{3}{2}}}\left[\psi'\psi\left(\frac{\delta U}{U}+\frac{\delta W}{W}\right)-2\psi\delta\psi'\right]-\frac{\al\vp'}{W}\delta\vp\right)\Omega_{(n-1)},
\end{equation}
here $\Omega_{(n-1)}$ denotes surface $n-1$--form. For nonminimally coupled part of the action we obtain:
\begin{equation}\label{var_nm}
(\delta Q-i_{\xi}\Theta)_{nm}=\frac{\eta(n-1)}{2}r^{n-2}\sqrt{\frac{U}{W}}\left(\frac{(\vp')^2}{2W^2}\delta W-\delta\left(\frac{(\vp')^2}{W}\right)+\frac{2\al r}{(n-1)\e}\vp'\delta\vp\right)\Omega_{(n-1)}.
\end{equation} 
Having combined latter two relations we can obtain total variation:
\begin{eqnarray}\label{var_tot}
\nonumber(\delta Q-i_{\xi}\Theta)_{tot}=r^{n-1}\sqrt{UW}\left((n-1)\left(1+\frac{\e(\vp')^2}{4W}\right)\frac{\delta W}{rW^2}+\frac{2}{UW}\frac{1}{\left(1-\frac{(\psi')^2}{\beta^2UW}\right)^{\frac{3}{2}}}\times\right.\\\left.\left[\psi'\psi\left(\frac{\delta U}{U}+\frac{\delta W}{W}\right)-2\psi\delta\psi'\right]-\frac{\eta(n-1)}{2rW}\delta\left(\frac{(\vp')^2}{W}\right)\right)\Omega_{(n-1)}.
\end{eqnarray}
We note that in the limit $\beta\rightarrow\infty$ the given above relation is reduced to the corresponding relation for standard linear Maxwell field \cite{Feng_PRD16}.  Using the written above relation (\ref{var_tot}) we can calculate total variation at the infinity as well as on the horizon. As a result at the infinity we write:
\begin{equation}\label{var_inf}
(\delta Q-i_{\xi}\Theta)_{tot}=((n-1)\delta\mu+4\psi_0\delta q)\Omega_{(n-1)}.
\end{equation}
The obtained relation is very simple and completely coincides with corresponding relation obtained for linear Maxwell field \cite{Feng_PRD16}. Taking into account relations for the total charge and gauge potential and performing integration over angular variables we can obtain relation for the variation of the gravitational Hamiltonian at the infinity $\delta{\cal H}_{\infty}$:
\begin{equation}\label{H_inf_calc}
\delta{\cal H}_{\infty}=\delta M-\Phi_q\delta Q
\end{equation}
where $\delta M$ is the variation of the black hole's mass which can be written in the form:
\begin{equation}\label{BH_mass}
M=\frac{(n-1)\omega_{n-1}}{16\pi}\mu
\end{equation}
It should be pointed out that for non-spherical topology of horizon the relation (\ref{H_inf_calc}) should be treated as the variation per unit volume. We would also like to emphasize that  obtained relation for the mass (\ref{BH_mass}) takes completely the same form as in case of standard General Relativity.

Taking the variation of the Hamiltonian at the horizon we arrive at the relation:
\begin{equation}\label{TD_diff}
\delta {\cal H}_{+}=\frac{(n-1)\omega_{n-1}}{16\pi}U\rq{}(r_+)r^{n-2}_{+}\delta r_{+}=\sqrt{U(r_+)W(r_+)}T\delta\left(\frac{{\cal A}}{4}\right)=\left(1+\frac{\e}{4}\frac{(\vp\rq{})^2}{W}\Big|_{r_+}\right)T\delta\left(\frac{{\cal A}}{4}\right).
\end{equation}
where ${\cal A}=\omega_{n-1}r^{n-1}_+$ is the horizon area of the black hole. Here we note that variation of the Hamiltonian at the horizon does not include a contribution from the gauge field due to the fact that  gauge potential equals to zero at the horizon. Latter relation for the variation of the Hamiltonian at the horizon can't be represented in the form $T\delta S$ which takes place in the standard General Relativity, in other words the form in the right hand side of the relation (\ref{TD_diff}) is not exact, this fact was noted in \cite{Feng_PRD16} and a specific procedure was proposed to derive a relation for black hole's entropy. To  write the first law of black hole mechanics it was proposed to introduce specific ``scalar charge'', related to the scalar field \cite{Feng_PRD16}. But as it was shown in our earlier works \cite{Stetsko_slr,Stetsko_PRD19} the ``scalar charges'' can be chosen in different way, and here we take the same form for them as in our previous works and rewrite the latter relation in the form:
\begin{equation}\label{var_horizon}
\delta{\cal H}_{+}=T\delta S+\Phi^{+}_{\vp}\delta Q^{+}_{\vp},
\end{equation}
where $S$ is the entropy of the black hole, $Q^{+}_{\vp}$ and $\Phi^{+}_{\vp}$ denote introduced ``scalar charge'' and related to it potential. These introduced values can be chosen in the form:
\begin{equation}\label{entropy}
S=\left(1+\frac{\e}{4}\frac{(\vp\rq{})^2}{W}\Big|_{r_+}\right)\frac{{\cal A}}{4},
\end{equation}
\begin{equation}\label{sc_pot}
Q^{+}_{\vp}=\omega_{n-1}\sqrt{1+\frac{\e}{4}\frac{(\vp\rq{})^2}{W}\Big|_{r_+}}, \quad \Phi^{+}_{\vp}=-\frac{{\cal A}T}{2\omega_{n-1}}\sqrt{1+\frac{\e}{4}\frac{(\vp\rq{})^2}{W}\Big|_{r_+}}.
\end{equation}
As we mentioned above the scalar ``charge'' $Q^{+}_{\vp}$ and conjugate potential $\Psi^{+}_{\vp}$ might be defined in other way, but in our case it allows to derive relation between the temperature and entropy from one side and introduced scalar ``charge'' and potential from the other one:  
\begin{equation}\label{rel_entr_pot}
\Phi^{+}_{\vp}Q^{+}_{\vp}=-\frac{{\cal A}T}{2}\left(1+\frac{\e}{4}\frac{(\vp\rq{})^2}{W}\Big|_{r_+}\right)=-2TS
\end{equation}

We point out here that contribution of scalar field charges was examinined for the first time in \cite{Gibbons_PRL96}. Using relations (\ref{variat_inf_hor}) and corresponding results for the variations at the horizon and at the infinity we can write the first law in the following form:
\begin{equation}\label{first_law}
\delta M=T\delta S+\Phi^{+}_{\vp}\delta Q^{+}_{\vp}+\Phi_{q}\delta Q
\end{equation}
The obtained relation (\ref{first_law}) has very simple form, similar to corresponding relation in standard General Relativity, but nevertheless the introduced definition of the entropy is not supported by some independent way of calculation. It should be pointed out that several attempts to calculate entropy with help of Euclidean methods were made, but  there they were related mainly to planar geometry \cite{Caceres_JHEP17} or chargeless case. It should be pointed out here that accurate application of Euclidean action requires properly regularized and renormalized action, this regularization means that we take into account Gibbons-Hawking-York boundary term (\ref{GHY_nm}) and renormalization or the regularized action should be performed to make the action finite, since we deal with asymptotically AdS solutions, but this issue is a subject of independent investigation and will be performed elsewhere.

\section{Conclusions}
In this work we consider particular case of general Horndeski gravity, namely the theory with nonminimal derivative coupling and we also take into account gauge filed minimally coupled just to gravity sector and given by a Lagrangian of Born-Infeld type. We obtain static solutions which represent black holes. Since the cosmological constant is taken into account it allowed us to consider not only spherically symmetric solution, but also to obtain topological solutions with nonspherical horizon, namely with flat ($\ve=0$) and hyperbolic ($\ve=-1$) ones. 

In general  the structure of the obtained solutions is complicated, but nevertheless they share some common features with black holes' solutions derived in the framework of General Relativity as well as in Horndeski Gravity. Firstly, all the obtained solutions have AdS-like behaviour at large distances, because of the presence of the cosmological constant $\L$ and due to the influence of the scalar field, but in contrast with the ordinary General Relativity, where the metric at large distances is completely defined by a term with bare cosmological constant $\Lambda$ in our case we have some effective constant which is defined by the bare one and coupling constants for the scalar field $\al$ and $\eta$. The behaviour of the metric for very small distances ($r\rightarrow 0$) depends on the type of topology as well as on the dimension of space. Namely, when $\ve\neq 0$ there are three different types of the behaviour of the metric function $U(r)$, for $n=3$, $n=4$ and $n\geqslant 5$, what is demonstrated by the relations (\ref{u_r0_3}), (\ref{u_r0_c4}) and (\ref{u_r0_c}) respectively, whereas for the flat horizon solution ($\ve=0$) the character of the function $U(r)$ for all $n\geqslant 3$ is the same (\ref{u_r0_f}). We also stress here that the leading terms for flat geometry ($\ve=0$) as well as nonflat geometry $\ve\neq 0$ if $n\geqslant 4$ is completely defined by the gauge field term and for the particular case $\ve\neq 0$ and $n=3$ the dominant term is of Schwarzschild type. In addition we would like to emphasize that the Kretschmann scalar (\ref{Kr_scalar}), which defines the character of singularity at the origin for the flat case $\ve=0$ and for the nonflat one $\ve\neq 0$ if $n\geqslant 5$ for small distances show the dependence $\sim 1/r^4$ and does not depend on the charge $q$, the same situation takes place for linear and power-law fields \cite{Stetsko_PRD19}. For $n=4$ and $\ve\neq 0$ the Kretshmann scalar has additional peculiarity of a logarithmic character (\ref{Kr_sc_4_c}) and for the case $n=3$ and $\ve\neq 0$ due to domination of the Schwarzschild term the Kretshmann scalar has completely different behaviour (\ref{Kr_sc_n3}).  We also obtained and examined the relations for the gauge field and gauge potential and it was shown that for $n=3$ and $\ve\neq 0$ the field and potential are nonsingular at the origin, whereas for other considered dimensions and types of geometry they are singular at this point. We point out here that in standard General Relativity with Born-Infeld field where the field and potential are nonsingular at all distances for all dimensions.

We have also examined some aspects of black hole's thermodynamics. First of all we have obtained relation for the temperature (\ref{temp_BI_gen}) and we show that in the limit $\b\rightarrow\infty$ we recover the relation (\ref{temp_lin_f}) which was derived in our previous work \cite{Stetsko_PRD19}. Careful analysis of the obtained relation shows that for large radius of the horizon $r_+$ the temperature increases almost linearly due to domination of AdS-term in this case. For small radius of horizon the situation is completely different, namely the leading term in this case is related to the gauge field, but it should be pointed out here that in case of finite $\b$ the character of dependence $T(r_+)$ is of the type $\sim \b r^{2-n}_+$, whereas for linear field one arrives at the asymptotic $\sim r^{3-2n}_+$  this difference can be explained by the fact that Born-Infeld gauge field has more moderate dependence of the $r_+$ than the linear one. To obtain the first law of black hole thermodynamics we have utilized Wald's approach which is applicable to general diffeomorphism-invariant  theories. Wald's method is well-posed, but nevertheless the definition of entropy is not an easy task, to introduce the entropy we followed the approach suggested in \cite{Feng_PRD16} and used in our work \cite{Stetsko_PRD19} where additional ``scalar charges'' were introduced. Here we point out that the choice we made to introduce ``scalar charge'' is not unique, in the paper \cite{Feng_PRD16} it was taken in a bit different form, this ambiguity and the fact that we do not have a corresponding ``charge'' related to the scalar field which appears due to integration of equations of motion makes this final step in Wald's procedure a bit unsatisfactory. To make this step well grounded we should have independent approach to define entropy and as a consequence to write the first law. As it is known the relation for entropy can be derived with help of Euclidean approach. In order to use Euclidean approach one should renormalize the total gravity action  (\ref{action}) to make it finite at infinity, we also point out here that several attempts to use Euclidean techniques were made, but they were mainly related to uncharged black hole \cite{Minamitsuji_PRD14} or black holes with flat topology of horizon \cite{Caceres_JHEP17}.
\section{Acknowledgments}
This work was partly supported by Project FF-83F (No. 0119U002203) from the Ministry of Education and Science of Ukraine.

\end{document}